\documentclass[11pt,a4paper]{article}
\usepackage[utf8]{inputenc}
\usepackage[T1]{fontenc}
\usepackage[english]{babel}
\usepackage{lmodern}
\usepackage{calc}
\usepackage{caption}
\usepackage{chngcntr}
\usepackage{enumitem}
\usepackage{setspace}
\usepackage{gensymb}
\usepackage{graphicx} 

\usepackage{authblk}

\usepackage{layout}
\usepackage{multicol,multirow}
\usepackage{amsmath,amsfonts,amssymb,amsthm}
\usepackage{mathrsfs}
\usepackage{float}

\usepackage[left=2.5cm,right=2.5cm,top=2.5cm,bottom=2.5cm]{geometry}
\usepackage[final]{pdfpages}
\usepackage{pstricks}
\usepackage{siunitx}

\usepackage{soul}
\usepackage{titlesec}
\usepackage{ulem}
\usepackage{url}

\usepackage{dsfont}
\usepackage{array}

\usepackage{wrapfig}
\usepackage{fancyhdr}
\usepackage[lastpage,user]{zref}

\usepackage{xcolor}
\usepackage{physics}
\usepackage{hyperref}
\usepackage{sectsty}
\definecolor{myblue}{rgb}{0,0.1,0.5}

\usepackage[style=numeric,sorting=none]{biblatex} %Imports biblatex package
\usepackage{color, colortbl}
\definecolor{lavendergray}{rgb}{0.77, 0.76, 0.82}
\definecolor{lightcornflowerblue}{rgb}{0.6, 0.81, 0.93}
\definecolor{flavescent}{rgb}{0.97, 0.91, 0.56}

\hypersetup{ 
colorlinks=true,
breaklinks=true, 
urlcolor=myblue,
linkcolor=black,
citecolor=myblue,
pdfauthor={Nicolas De Angelis}
}

\usepackage{varwidth}
\DeclareCaptionFormat{myformat}{
  \begin{varwidth}{\linewidth}
    \centering
    #1#2#3
  \end{varwidth}
}

\graphicspath{{./Pictures/}}

\begin{document}

\title{Characterisation of the MUSIC ASIC for large-area silicon photomultipliers for gamma-ray astronomy}

\author[a]{Nicolas De Angelis\footnote{Corresponding author.\newline \href{mailto:nicolas.deangelis@unige.ch}{nicolas.deangelis@unige.ch}}} %0000-0002-2498-0213
\author[b]{David Gascón} %0000-0001-9607-6154
\author[c,b]{Sergio Gómez} %0000-0002-3064-9834
\author[a]{Matthieu Heller} %0000-0003-1215-0148
\author[a]{Teresa Montaruli} %0000-0001-5014-2152
\author[a]{Andrii Nagai} %0000-0002-7150-4527

\affil[a]{DPNC, Université de Genève, 24 Quai Ernest Ansermet, CH-1205 Geneva, Switzerland}
\affil[b]{Institute of Cosmos Sciences - University of Barcelona (ICC-UB), Martí i Franquès, 1, 08028, Barcelona, Spain}
\affil[c]{Serra Húnter Fellow, Polytechnic University of Catalonia (UPC), Eduard Maristany, 16, 08019, Barcelona, Spain}

\renewcommand{\thefootnote}{\arabic{footnote}}

\date{\today}

\maketitle

\section*{Abstract}

Large-area silicon photomultipliers (SiPMs) are desired in many applications where large surfaces have to be covered. For instance, a large area SiPM has been developed by Hamamatsu Photonics in collaboration with the University of Geneva, to equip gamma-ray cameras employed in imaging atmospheric Cherenkov telescopes. 
The sensor being about 1~cm$^2$, a suitable preamplification electronics has been investigated in this work, which can deal with long pulses induced by the large capacitance of the sensor. The so-called Multiple Use SiPM Integrated Circuit (MUSIC), developed by the ICCUB (University of Barcelona), is investigated as a potential front-end ASIC, suitable to cover large area photodetection planes of gamma-ray telescopes.
The ASIC offers an interesting pole-zero cancellation (PZC) that allows dealing with long SiPM signals, the feature of active summation of up to 8 input channels into a single differential output and it can offer a solution for reducing power consumption compared to discrete solutions. Measurements and simulations of MUSIC coupled to two SiPMs developed by Hamamatsu are considered and the ASIC response is characterized.\\
%The 5\textsuperscript{th} generation sensor of the Low Cross Talk technology coupled to MUSIC turns out to be a good solution for gamma-ray cameras.\\

\noindent\textit{Keywords:} Photon detectors for UV, visible and IR photons (solid-state) (PIN diodes, APDs, Si-PMTs, G-APDs, CCDs, EBCCDs, EMCCDs, CMOS imagers, etc); Front-end electronics for detector readout; Cherenkov detectors
%\noindent\textit{Keywords:} SiPM readout, Silicon PhotoMultiplier, photo-detection, photo-sensors, ASIC, pole-zero cancellation, Cherenkov telescope, gamma-ray astronomy.

\newpage
\tableofcontents

\newpage
\section{Introduction: Current challenges of SiPMs in gamma-ray astronomy}
\label{sec:intro}

The next generation of cameras for gamma-ray astronomy are all moving toward silicon photo-multipliers (SiPMs) \cite{2019NIMPA.926...36K,ambrosi2022,key-Camera,2020PMB....65qTR01G,key-Sensors,Montaruli:2020vjj} for their stability, robustness and higher sensitivity compared to photomultiplier tubes (PMTs).
Additionally, SiPMs are more and more evolving toward tailored technologies for a given application. For instance, custom size SiPMs can be used to read out fiber ribbon (e.g. LHCb SciFi tracker \cite{LHCb_SciFi}, HERD \cite{HERD_ribbon}) or build pixels of a specific size (e.g. LHAASO WFCTA cameras \cite{LHAASO:2020unt}). Because each SiPM technology, micro-cell size or even sensor size cannot be produced and tested in the laboratory, it is very important to be able to foresee by means of simulations the response of the readout chain to a change in sensor type. This is even more true when the sensor has a custom size or shape~\cite{hex_sensor, LHAASO:2020unt} which requires most of the time a custom photo-mask and therefore prohibitive cost when just a trial sample is needed.

Our work was triggered by the needs of the SST-1M camera, described in \cite{key-Camera}, though it can be generalized\footnote{In order to so, the research team has to have a clear set of performance requirements and access to the simulation model of the readout electronics and to the sensor Corsi parameters. Both information can be obtained from the sensor and the ASIC manufacturers.}. The SST-1M camera is a SiPM-based fully digitizing camera, developed for Cherenkov imaging.
%and is a perfect example of this problem. 
It was originally developed in the frame of the Cherenkov Telescope Array (CTA) project~\cite{CTAO}. 
It is composed of a photo-detection plane (PDP), responsible for detecting Cherenkov light and pre-amplifying the signal, and of a digitizing system (DigiCam) that applies a trigger selection and digitizes out the signal to the camera server. 
The entrance of the camera is protected by a filtering window to avoid dust from entering the PDP and to filter background-dominated wavelengths\footnote{A typical Cherenkov photon has a wavelength from near-UV up to $\sim$550~nm. Cutting off high wavelengths (above 540~nm for the entrance window) allows for reducing the night-sky background (NSB), whose main emission is above 600~nm}. 
The PDP is based on hexagonal SiPMs coupled to conically-shaped hollow light guides. 
The SiPMs are read out by a two stages front-end electronics (FEE) \cite{key-Electronics}. 
The FEE consists of a preamplification board and a slow control board that routes the analog signals from the preamplification stage to DigiCam. An exploded view of both the camera and a single PDP module is shown in Figure \ref{fig:sst1m_camera}.

\begin{center}
\begin{figure}[H]
\captionsetup{format=myformat}
\begin{centering}
\includegraphics[height=.36\textwidth]{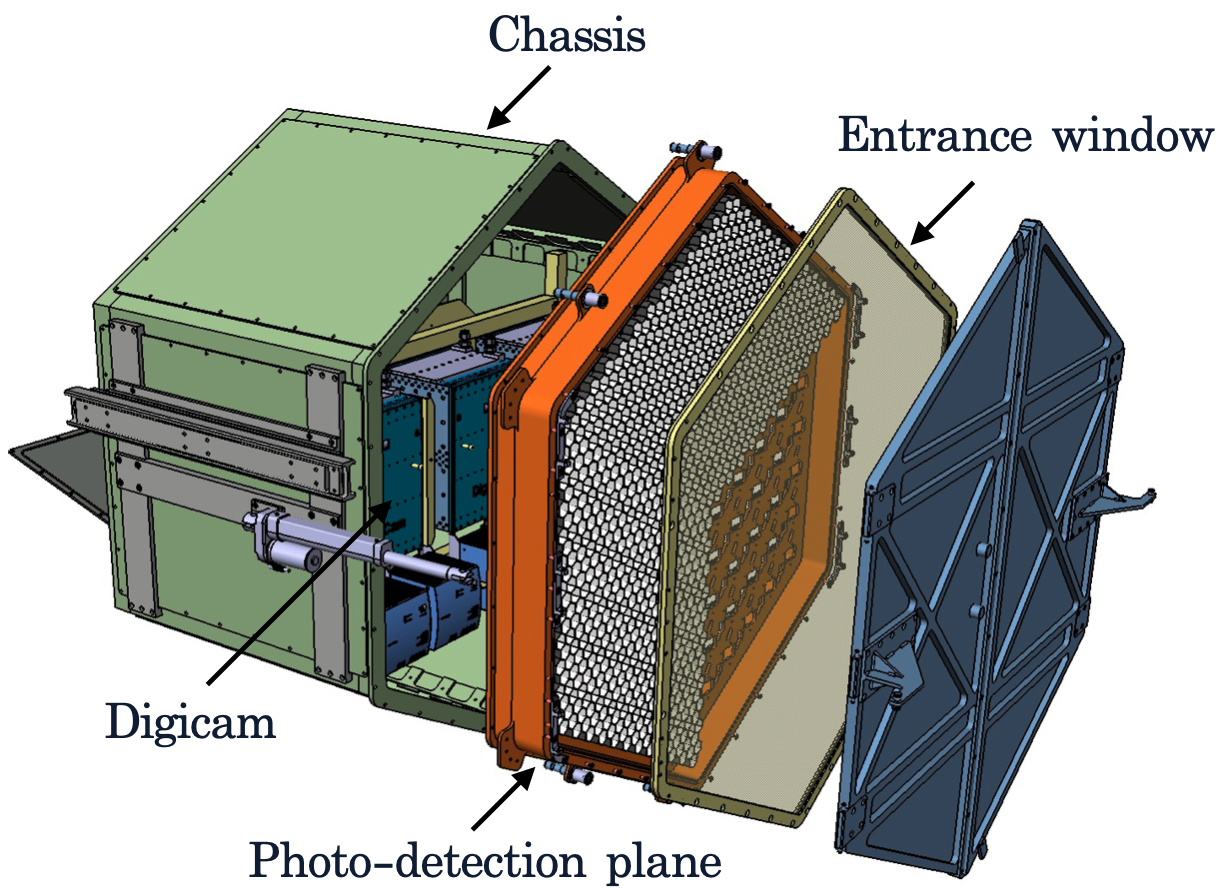}\includegraphics[height=.36\textwidth]{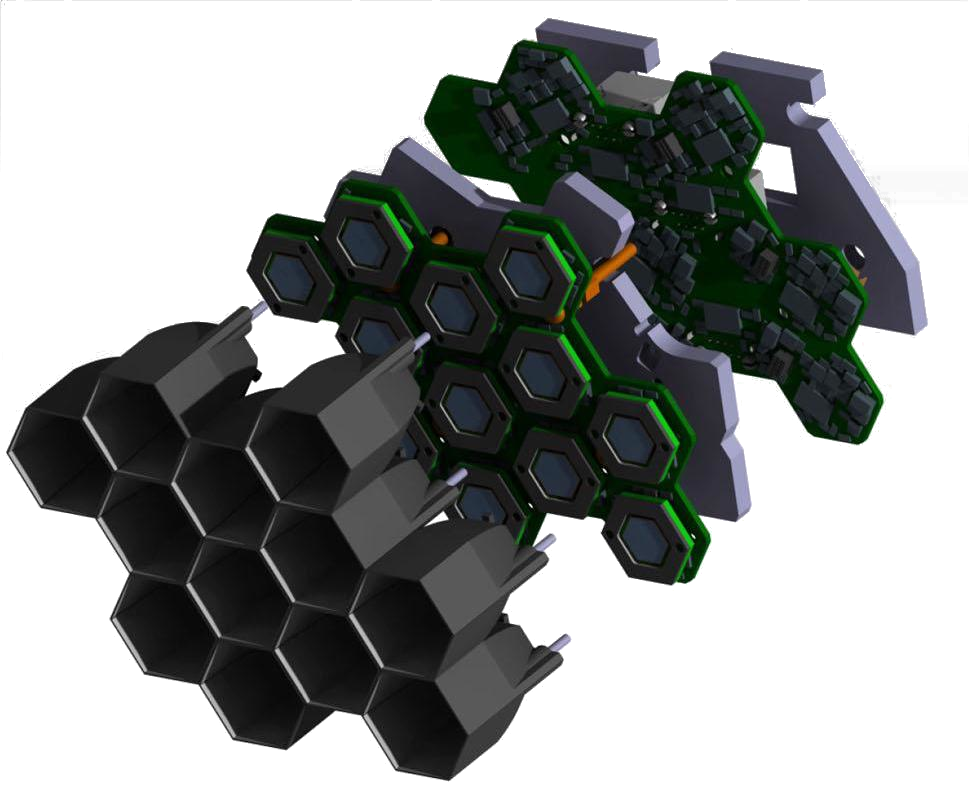}
  \par\end{centering}
  \protect\caption{Exploded CAD view of the SST-1M camera and of a PDP module \cite{key-Camera}.}
    \label{fig:sst1m_camera}
\end{figure}
\par\end{center}

The overall optical efficiency of the camera \cite{key-Camera} (conversion factor from photons to photoelectrons) was found to be 18.7\%. In order to increase such value to more than 20\%, as the light guide design and the window transmittance are already optimized for high efficiency, we investigate an upgrade of the sensor to a higher photo-detection efficiency (PDE). 
Two sensors from Hamamatsu (HPK) were originally considered for this upgrade: the so-called LCT5 sensor (S13360), based on the same low cross-talk technology as the already adopted LCT2 sensor in the current camera, and the LVR3 (S14520), based on the low voltage resistor technology.  
The results are compared with the LCT2 sensor, with which the current camera is instrumented. The results are presented in Figure~\ref{fig:SiPMMeasurements}. The measurements and analysis procedure are described in \cite{key-Sensors, key-SENSE}.  
The LCT5 sensor, operated at the same optical crosstalk value as the LCT2 of 8\%, reaches 25.4\% overall optical efficiency, while LVR3 reaches 28.0\%.
These two sensors not only have a higher PDE but also a higher capacitance and quenching resistance, inducing long-tail signals. The LVR3 has a 25\% larger diode capacitance and 200\% higher quenching capacitance which translates into slower response time and a larger electronic noise.
As a result, the width of the signal integration window has to be larger for the LVR3 than for the LCT5 resulting in higher contamination from NSB photons hence a worse charge resolution. In addition, signals from NSB photons will more easily pile up increasing the probability to trigger on noise for a given threshold. The consequence would therefore be an increase in the energy threshold. 
Therefore, we decided to select the LCT5 device for this study. This decision is even more motivated when foreseeing their use in even larger telescopes where pulses as short as 3~ns full width at half maximum (FWHM) are targeted \cite{CTALSTProject:2021imy}. 

\begin{center}
\begin{figure}[H]
\captionsetup{format=myformat}
\begin{centering}
\includegraphics[height=.45\textwidth]{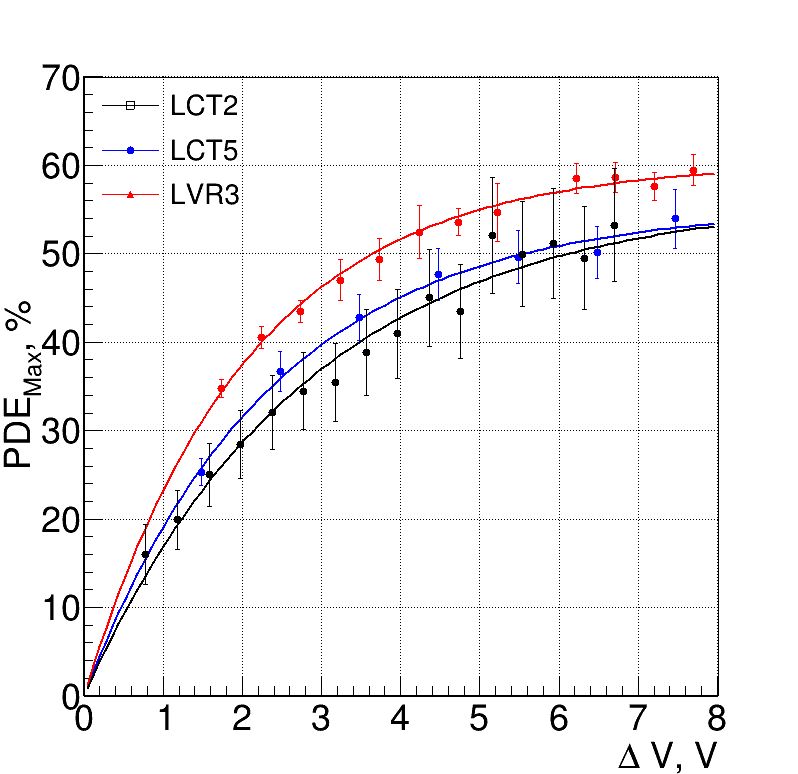}\includegraphics[height=.45\textwidth]{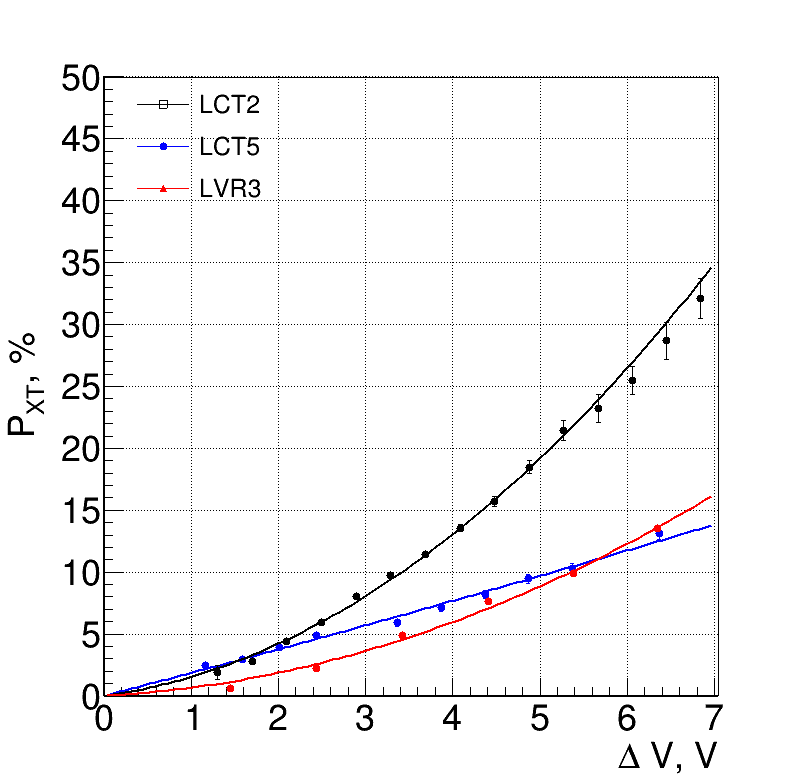}
  \par\end{centering}
  \protect\caption{Comparison of SIPM peak $PDE$ (left) and crosstalk probability $P_{XT}$ (right) as a function of over-voltage $\Delta V$.}
    \label{fig:SiPMMeasurements}
\end{figure}
\par\end{center}

Fine-tuning of the current analog front-end electronics of the SST-1M camera, in order to get sharp enough signals, is not sufficient. 
Hence, in order to profit from the new sensors we describe the redesign of the preamplification stage of the front-end electronics that can deal with these higher-capacitance SiPMs.

Given these constraints, the only ASIC at the time of this study that allows coping with the large sensor capacitance is the Multiple Use SiPM Integrated Circuit (MUSIC) ASIC~\cite{key-MUSIC_paper, key-MUSIC_journal}. This 8-channel pre-amplifying ASIC has been developed at the University of Barcelona. 
It contains several functionalities: (1) summation of up to eight input channels and providing a differential output; (2) individual readout of the time over threshold response, and (3) individual analog response using a single-ended output driver.
Both output drivers are capable of driving 50~$\Omega$. Each operating mode, summation and single input processing includes a configurable dual-gain option. Moreover, the summation mode implements a dual-gain output providing four different gains for this operating mode. 
Figure \ref{fig:sim_music} shows the architecture and the main functionalities of the MUSIC ASIC.

\begin{center}
\begin{figure}[H]
\captionsetup{format=myformat}
\begin{centering}
\includegraphics[height=.5\textwidth]{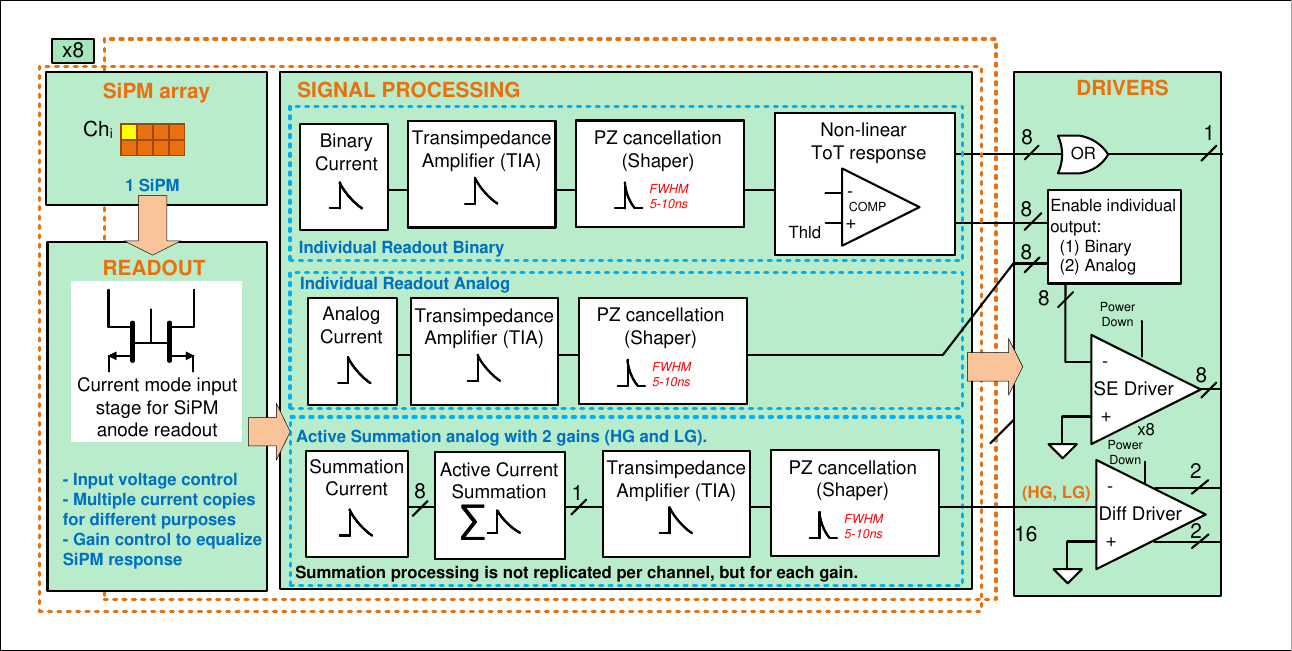}
  \par\end{centering}
  \protect\caption{MUSIC ASIC architecture illustrating the different operating modes.~\cite{key-MUSIC_journal}.}
    \label{fig:sim_music}
\end{figure}
\par\end{center}

The ASIC also offers the possibility to apply a pole-zero cancellation in any operating mode, reducing the SiPM pulse FWMH. For the particular case studied here, this allows reducing the signal tail and summing from four channels of a pixel as it is done in the SST-1M front-end electronics. 
This is particularly interesting for the current SiPM technologies which exhibit large capacitance. Furthermore, this ASIC would also reduce the power consumption of the FEE. 
The idea would be to use a MUSIC board per pixel in high gain (HG) mode to readout, sum, and amplify the signals of the 4 channels of a pixel and to send a single signal to the slow control board. The integration of the MUSIC ASIC in the current camera topology is shown in figure \ref{fig:music_integration_camera}.

\begin{center}
\begin{figure}[H]
\captionsetup{format=myformat}
\begin{centering}
\includegraphics[width=.8\textwidth]{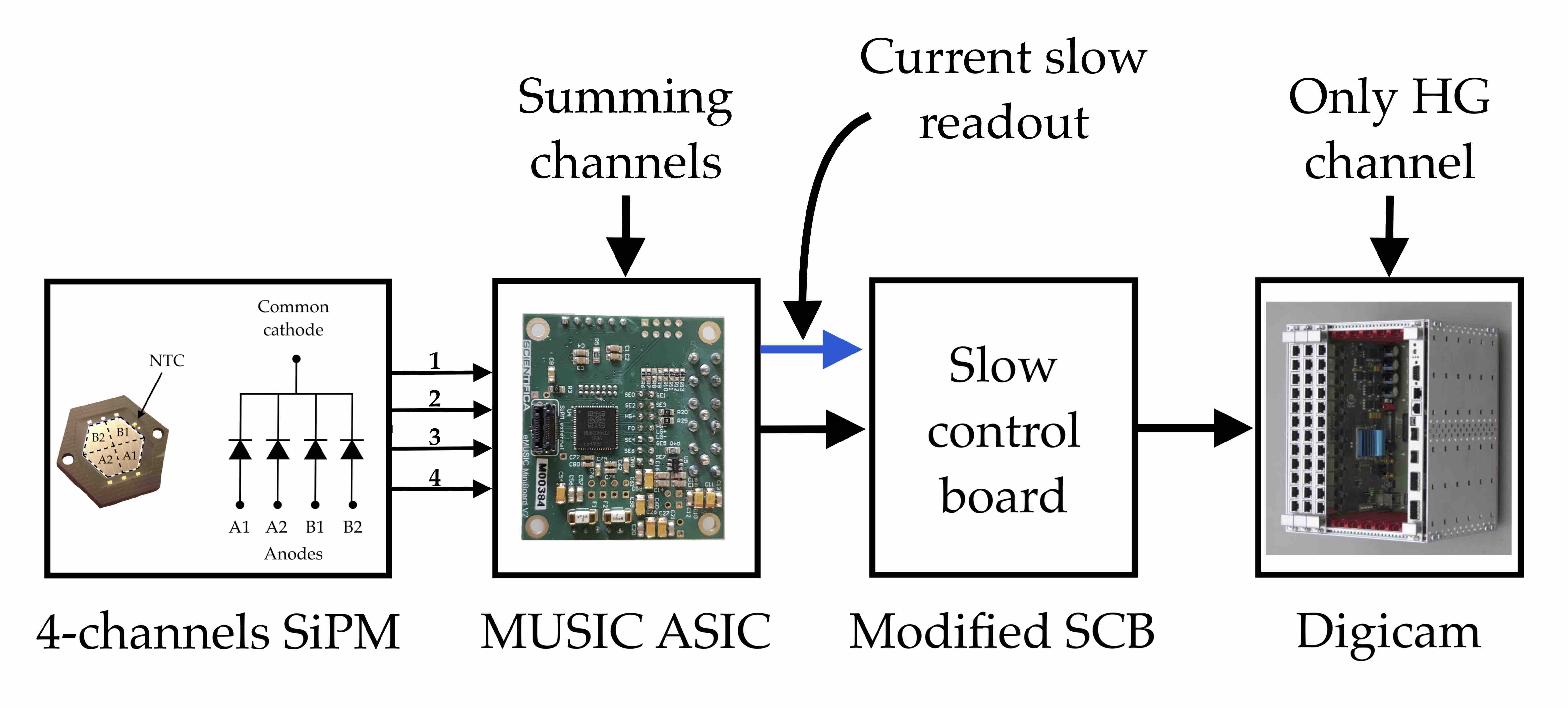}
  \par\end{centering}
  \protect\caption{MUSIC ASIC as preamplification electronics for the SST-1M camera \cite{key-thesisNDA}.}
    \label{fig:music_integration_camera}
\end{figure}
\par\end{center}

%The measurement setup used to characterize the MUSIC ASIC with the LCT5 sensor is first described, followed by the resulting pulse shapes and photo-electron spectra obtained by coupling the MUSIC ASIC with SiPMs of area 3$\times$3~mm$^2$. A study of the linearity of the HG channel is then presented. Finally, the measured pulses are compared with the simulated MUSIC response. The study is completed by the simulation of the LCT5 sensor in hexagonal shape enabling performance extrapolation for future cameras.

The presentation of the work is structured as follow:
\begin{itemize}
\item Description of the measurement and simulation setups used to characterized the MUSIC ASIC with the LCT5 sensor.
\item Pulse shape and photo-electron spectra measurement results obtained by coupling the MUSIC ASIC with a 3$\times$3~mm$^2$ SiPM and study of the linearity of the HG channel response.
\item Pulse shape comparison between measurements and simulated predictions for different PZC configurations.
\item Response prediction of the MUSIC ASIC for an hexagonal LCT5 sensor, enabling performance extrapolation for future cameras with large area custom sensors.
\end{itemize}

\section{Experimental and simulation setups}
\label{sec:setups}

\begin{center}
\begin{figure}[H]
\captionsetup{format=myformat}
\begin{centering}
\includegraphics[width=\textwidth]{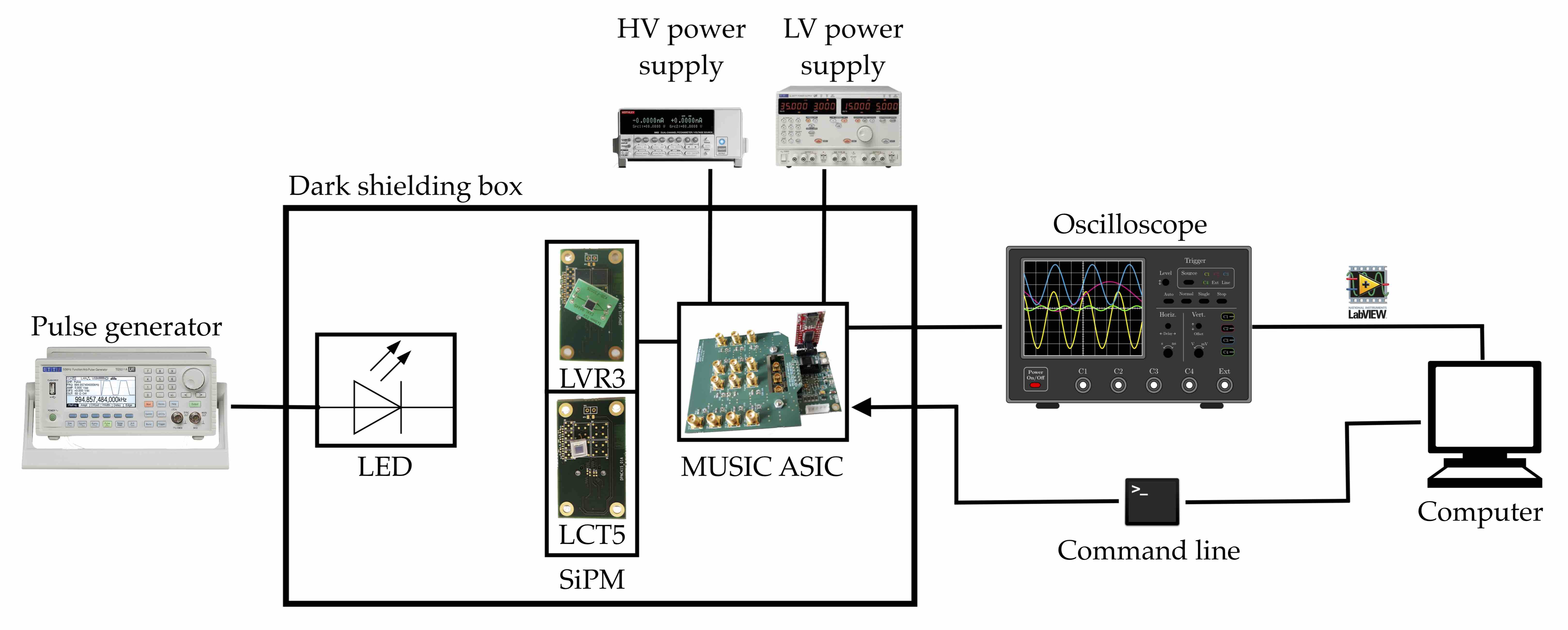}
  \par\end{centering}
  \protect\caption{Schematic view of the experimental setup~\cite{key-thesisNDA}.}
    \label{fig:music_setup}
\end{figure}
\par\end{center}

The measurement setup used to characterize the LCT5 sensor together with the MUSIC is shown in Figure~\ref{fig:music_setup}. Inside the dark box, an LCT5 sensor is soldered on a small dedicated PCB connected to the MUSIC board. Two other boards are connected to the MUSIC PCB, one to communicate with the computer and configure the MUSIC ASIC, and the other to readout the single-ended and differential outputs using SMA cables. An LED with a wavelength of 525~nm is placed in the dark box and connected to a pulse generator placed outside the box in order to control the light source. The single-ended and differential channels are read out using a Teledyne Lecroy 620Zi oscilloscope. The oscilloscope is controlled by the computer via a LabView\textsuperscript{TM} script in order to automate the data acquisition. Two power supplies are connected to the MUSIC board: the High Voltage power supply to bias the sensors ($\sim$40-60~V) and the Low Voltage power supply in order to bias the MUSIC (6.5~V). The instruments used in this setup are listed in Table \ref{tab:instrument_list}.

\begin{center}
\begin{figure}[H]
\captionsetup{format=myformat}
\begin{centering}
\includegraphics[height=.35\textwidth]{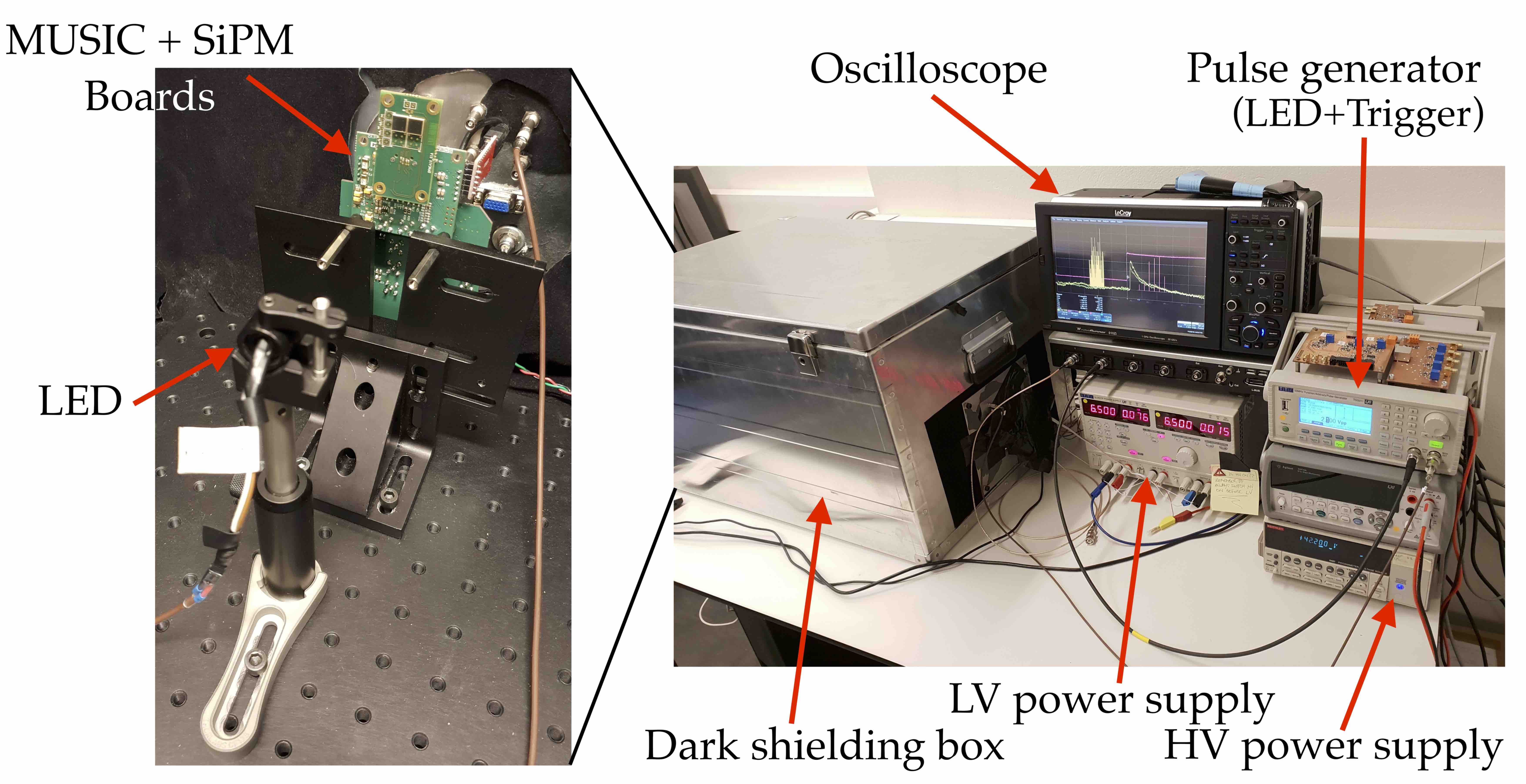}\includegraphics[height=.35\textwidth]{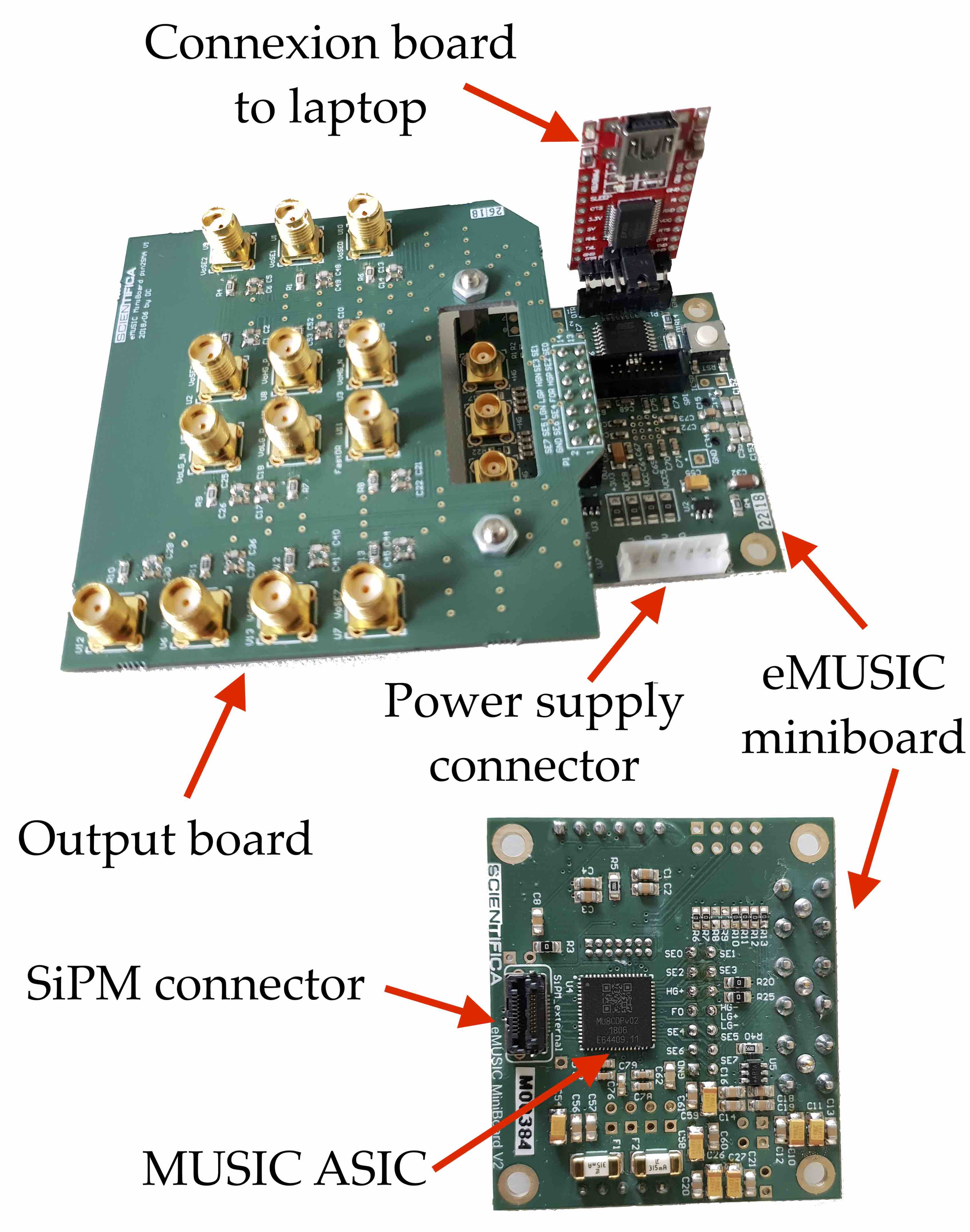}
  \par\end{centering}
  \protect\caption{Picture of the experimental setup \cite{key-thesisNDA}}
    \label{fig:music_setup_pic}
\end{figure}
\par\end{center}

\begin{table}[ht]
    \small
	\centering
	\renewcommand*{\arraystretch}{1.4}
	\setlength\arrayrulewidth{1.5pt}
	\begin{tabular}{|l|c|r|}
		\hline
		\rowcolor{flavescent} 
		\textbf{Function}  & \textbf{Instrument} & \textbf{Model}  \\
		\hline
		\rowcolor{lightcornflowerblue}
		SiPM bias & Sourcemeter & Keithley 2400 \\	 
		\hline
		\rowcolor{lavendergray} 	
     		Low power supply & Power Supply & Aim-TTi QL355TP \\
		\hline
		\rowcolor{lightcornflowerblue}
     		LED pulser & Arbitrary waveform generator & Aim-TTi TG5012 \\ 
		\hline
		\rowcolor{lavendergray} 	
     		Pre-amplifier & ASIC demo board & eMUSIC \\ 
		\hline
		\rowcolor{lightcornflowerblue}
     		Waveform acquisition & Oscilloscope & Teledyne Lecroy 620Zi\\
		\hline
		\rowcolor{lavendergray} 	
     		Control software & LabView &  \\ 
		\hline
	\end{tabular}
	\caption{List of instruments and software used for the SiPM characterization setup}

	\label{tab:instrument_list}

\end{table}

A picture of the measurement setup is shown in Figure~\ref{fig:music_setup_pic}. The light source excitation pulse is set with the shortest possible width (20~ns) and a frequency of 1~kHz. 
The MUSIC is configured via the command line tool of the computer with a windows executable specifying the channels that have to be enabled both in single-ended and differential modes. 
The internal gain of the analog processing chain can also be tuned by configuring an internal register. Finally, the PZC can be enabled and tuned using two parameters R (3-bits) and C (5-bits) that adjust the values of the resistors and capacitor in the shaper circuit \cite{key-MUSIC_journal}.

\begin{center}
\begin{figure}[H]
\captionsetup{format=myformat}
\begin{centering}
\includegraphics[height=.5\textwidth]{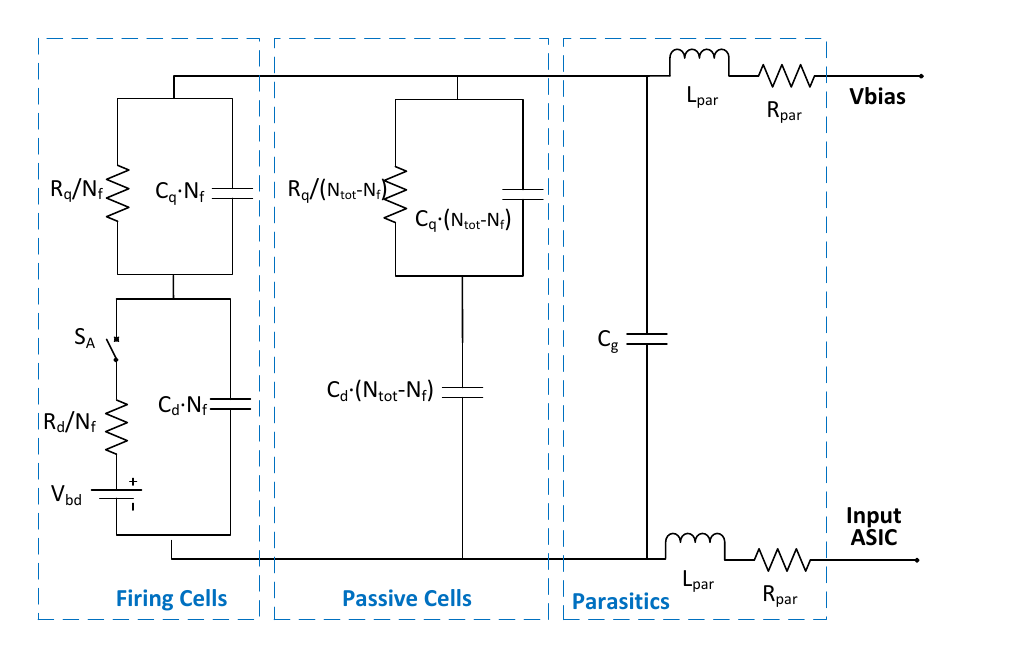}
  \par\end{centering}
  \protect\caption{SiPM model employed in the simulations.}
    \label{fig:sim_sipm}
\end{figure}
\par\end{center}

Simulations have been performed using the Cadence Design environment with Spectre simulator\footnote{Cadence Design Systems, Inc. \url{https://www.cadence.com/}} in order to compare to experimental results. The simulations have been performed with both LCT5 and LVR3 technologies, the former to compare with experimental measurements and to extrapolate the behavior to larger sensors, and the latter to address the concerns about the SNR expressed in Section~\ref{sec:intro}. Each SiPM is simulated following the circuit structure depicted in Figure~\ref{fig:sim_sipm}.

$R_{q}$ and $C_{q}$ are the quenching resistor and capacitance respectively, $R_{d}$ and $C_{d}$ are the diode resistance and capacitance, $C_{g}$ the parasitic capacitance, $N_{tot}$ is the number of micro-cells, $N_{f}$ the number of fired cells and $V_{bd}$ is the breakdown voltage of the SiPM. 
$S_{A}$ is a switch implemented in verilog-A code that models the arrival time of the photons. A complete description of the SiPM circuit model can be found in \cite{key-simus-paper}. 
In this simulation, the SiPM is connected to the bias voltage from one side and is DC coupled to the input of the MUSIC chip from the other side. 

It is important to highlight that the simulations have been done at the transistor level by using the MUSIC circuit. 
Being able to perform full ASIC post-layout simulations is much more accurate than using the simulation models that are commonly made available by major manufacturers (e.g. Texas Instruments, Analog devices, etc.). Note that the combination of both the MUSIC and the SiPM has been simulated for different settings of the ASIC.

The simulations have been performed without considering the RC parasitic of the layout in order to compute the SNR, but the pulse shape analysis has been done using the parasitics of the layout. We used the schematic of the full circuit connected to a SiPM modeled using the Corsi model detailed in Figure~\ref{fig:sim_sipm}. In particular, the anode of the sensor is connected to the input of the ASIC, just by adding an inductance and a resistance in series (1~nH and 3~$\Omega$ respectively) and a capacitor connected to ground to model the resistance and inductance of the connection and the capacitance of the pad ($\sim$5~pF). The complete parasitic of the circuit has not been used in order to accelerate transient simulations. As the goal of this study was to compare the performance of the ASIC, between the different configurations of the sensor, the inaccuracies of the simulations affect all of the sensor configurations equally. 
%Besides, the experimental measurements were used to crosscheck that we can actually identify and distinguish the different photons with the real system, i.e., that we have a high enough SNR to identify single photons.

A simplified schematic model used for the simulations is illustrated in Figure \ref{fig:simulation_setup}. Note that the circuit of the SiPM model and the architecture of the ASIC are detailed in Figures \ref{fig:sim_sipm} and \ref{fig:sim_music}, respectively.

\begin{center}
\begin{figure}[H]
\captionsetup{format=myformat}
\begin{centering}
\includegraphics[width=.7\textwidth]{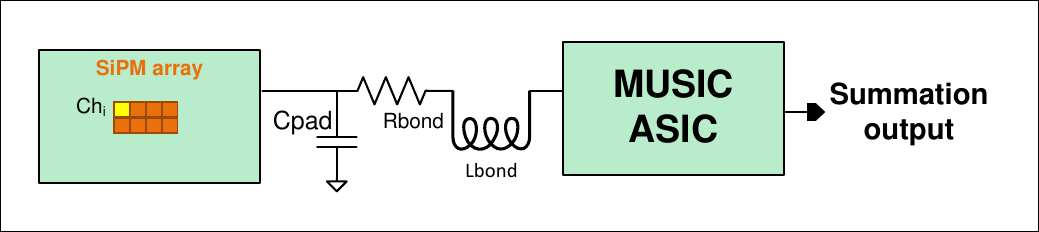}
  \par\end{centering}
  \protect\caption{Simplified schematic model of the MUSIC ASIC connected to a SiPM array used for the simulations.}
    \label{fig:simulation_setup}
\end{figure}
\par\end{center}

%\newpage
\section{LCT5 pulse shapes, photoelectron spectra, and linearity measurements with the MUSIC high gain channel}

The LCT5 technology has been measured with all possible PZC configurations. Figure \ref{fig:pulse_lct5_before_after_pzc} shows the measured pulse shape of a 3$\times$3~mm$^2$ LCT5 sensor with different R and C configurations.

\begin{center}
\begin{figure}[H]
\captionsetup{format=myformat}
\begin{centering}
\includegraphics[height=.42\textwidth]{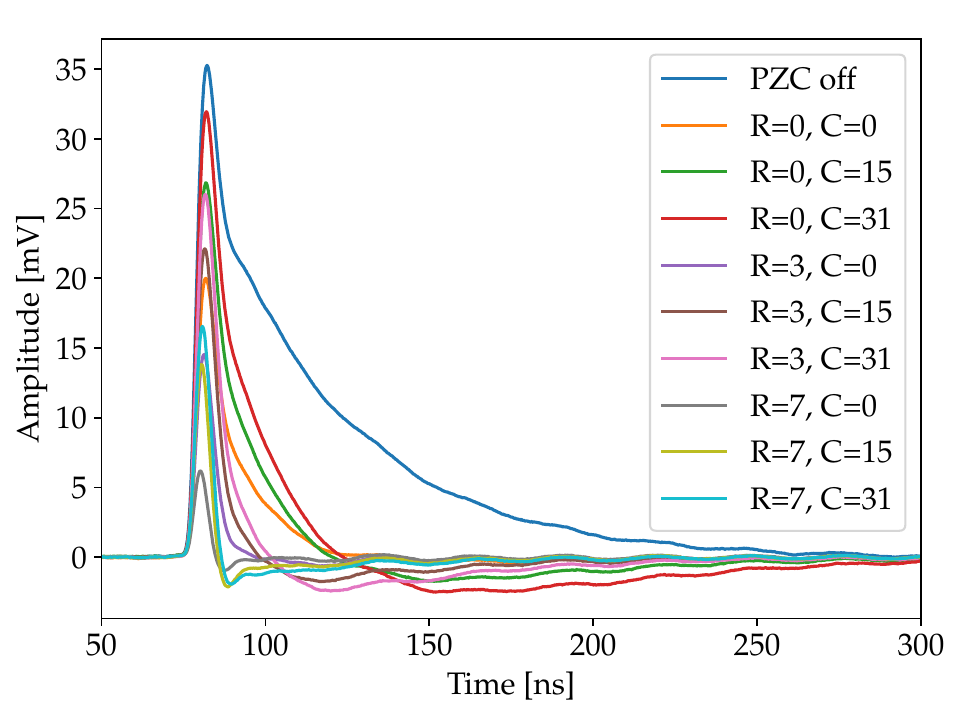}
  \par\end{centering}
  \protect\caption{Pulse shape for the 3$\times$3~mm\textsuperscript{2} LCT5 sensor at 4.4~V over-voltage for different PZC configurations compared to the pulse without cancellation}
    \label{fig:pulse_lct5_before_after_pzc}
\end{figure}
\par\end{center}

The initial pulse has a length of 120~ns, while applying a PZC allows to shorten it below 40~ns even reaching a recovery time smaller than 10~ns, thus satisfying the required integration time (see Figure~\ref{fig:recovery_pzc_lct}).
As visible in Figure~\ref{fig:pulse_lct5_before_after_pzc}, the higher is the PZ cancellation, the smaller is the pulse amplitude, which results in a worse signal-to-noise ratio. One solution to compensate for it is to increase the over-voltage at the cost of higher correlated and uncorrelated noises. 
For each configuration, the characteristics of the sensor and readout chain are derived from the generalized Poisson fit of the multiple photo-electron spectrum~\cite{Alispach_2020}.
As described in \cite{key-MUSIC_journal}, measurements performed using the 6$\times$6~mm$^2$ LCT5 sensor show that it is possible to count photons even with this larger area sensor. More measurements with the LCT5 technology will be detailed in section \ref{sec-hexa}. The LCT5 coupled to MUSIC could therefore be a suitable solution for sensor/electronics upgrade.

% Removed: The measurements have been performed with the 3$\times$3~mm$^2$ LCT5 sensor since it was the only size of LCT5 that we had. 

\begin{center}
\begin{figure}[H]
\captionsetup{format=myformat}
\begin{centering}
\includegraphics[height=.4\textwidth]{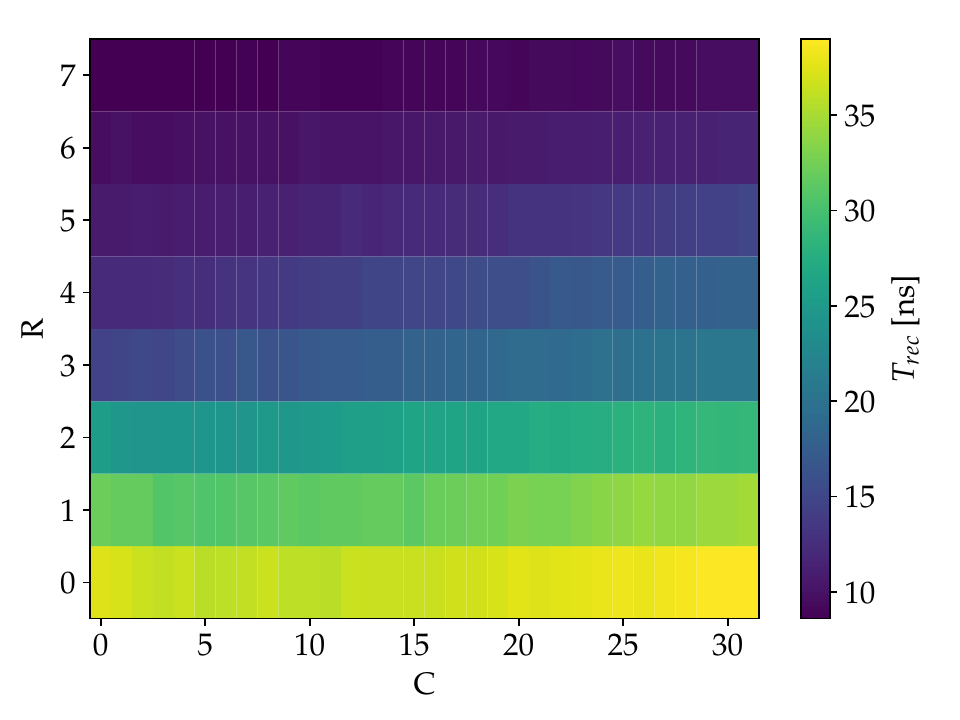}\includegraphics[height=.4\textwidth]{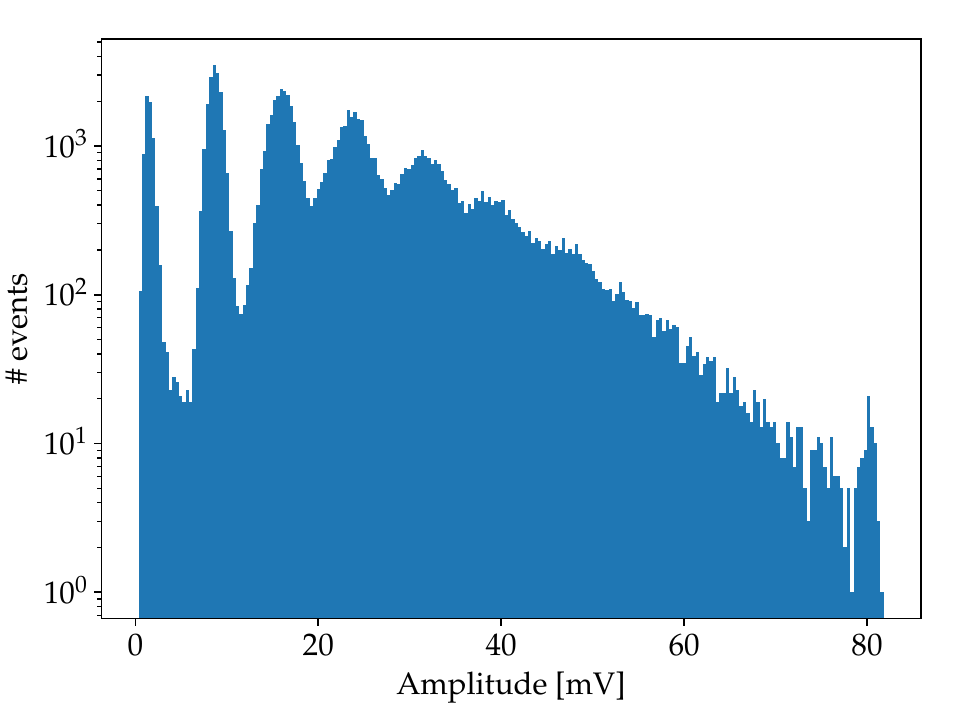}
  \par\end{centering}
  \protect\caption{Recovery time on the z-axis as a function of R and C for the LCT5 pulse and for each PZC configuration (left, adapted from \cite{key-thesisNDA}) and photo-electron spectrum for the 3$\times$3~mm\textsuperscript{2} LCT5 sensor at 4.4~V over-voltage (right, \cite{key-thesisNDA}).}
    \label{fig:recovery_pzc_lct}
\end{figure}
\par\end{center}

% VOY POR AQUI

%\section{Measurement of linearity of the high gain channel with the LCT5 sensor}

The evolution of the pulse shape as a function of the light level has also been measured in order to study the behavior in saturation of the LCT5 3$\times$3~mm$^2$ sensor coupled to MUSIC. 
The differential outputs are used for these measurements. The idea is to see if the HG channel has a predictive behavior in saturation in order to use it to reconstruct signals up to a few thousand of photons with a good resolution. 
Indeed, for cameras that would already adopt a single gain channel, as it is for DigiCam \cite{key-Electronics}, or for cameras with a large amount of pixels with demanding power-consumption constraints, doubling the number of channels, using both HG and LG outputs, may be unfeasible. 
%Increasing the number of channels would increase the power consumption and the price of the whole system. 
Figure~\ref{fig:light_level_scan_pulse_shapes} shows the evolution of the LCT5 pulse shapes with the bias voltage of the LED in HG mode. The pulses are normalised to the maximum amplitude in order to bring out the shape evolution. As expected, even when the amplitude saturation is reached,  the charge, and therefore, the pulse duration, are still increasing. 
As a matter of fact, the SiPM amplitude and charge increase linearly with the number of photons detected~\cite{Collazuol2019}. When the amplitude saturates, the number of detected photons can only be inferred from the charge of the signal. In such case, the integration window used to extract the charge has to adapt to the signal width. In this study, the integration window is defined by the rising and falling edges for a threshold of 5\% of the maximum amplitude.

\begin{center}
\begin{figure}[H]
\captionsetup{format=myformat}
\begin{centering}
\includegraphics[height=.4\textwidth]{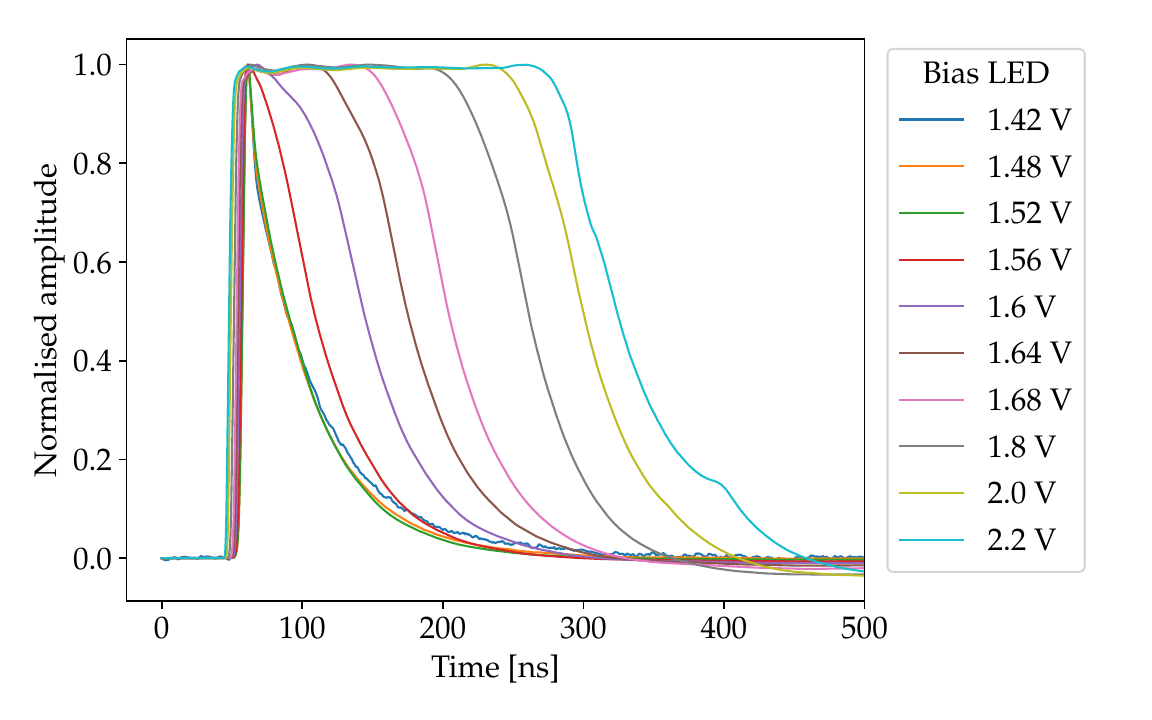}
  \par\end{centering}
  \protect\caption{Pulse shapes of the LCT5 sensor as a function of light level for the HG differential output \cite{key-thesisNDA} without PZC.}
    \label{fig:light_level_scan_pulse_shapes}
\end{figure}
\par\end{center}

\vspace{-.5cm}
The amplitude and the charge are extracted from the previous pulses and are plotted as a function of the bias voltage of the LED for the HG channel. This is shown in figure \ref{fig:amp_charge_nocalib_calibLED}. In order to describe the saturation using meaningful physical quantities, the LED needs to be calibrated, i.e. the amount of light emitted by the LED, and later detected by the SiPM, should be characterized as a function of the bias voltage applied to the LED.
\vspace{-.2cm}

\begin{center}
\begin{figure}[H]
\captionsetup{format=myformat}
\begin{centering}
\includegraphics[height=.4\textwidth]{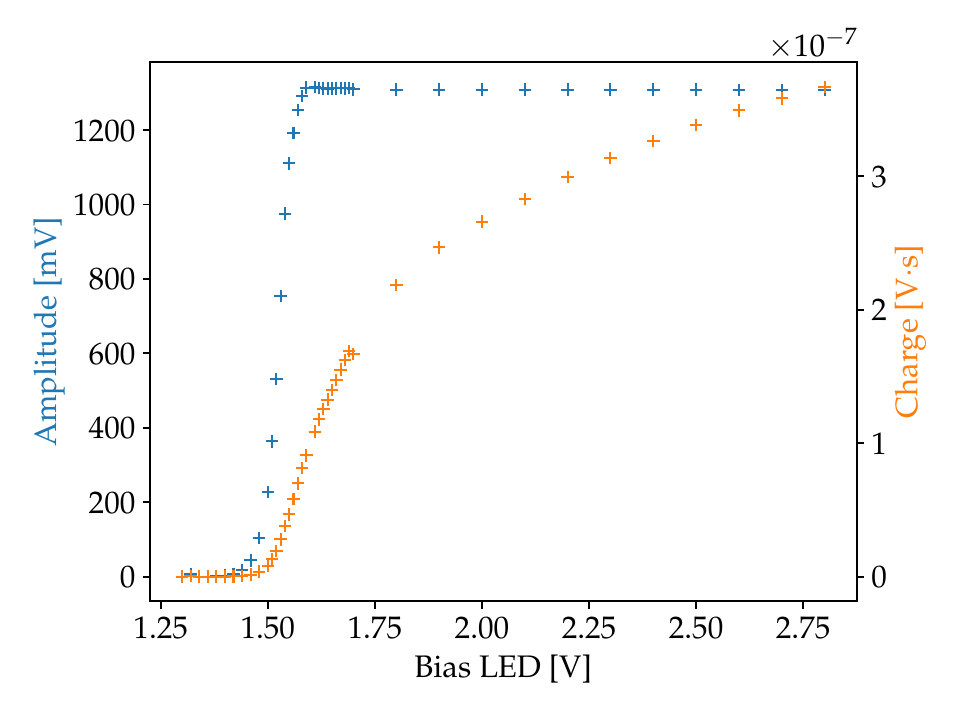}\includegraphics[height=.4\textwidth]{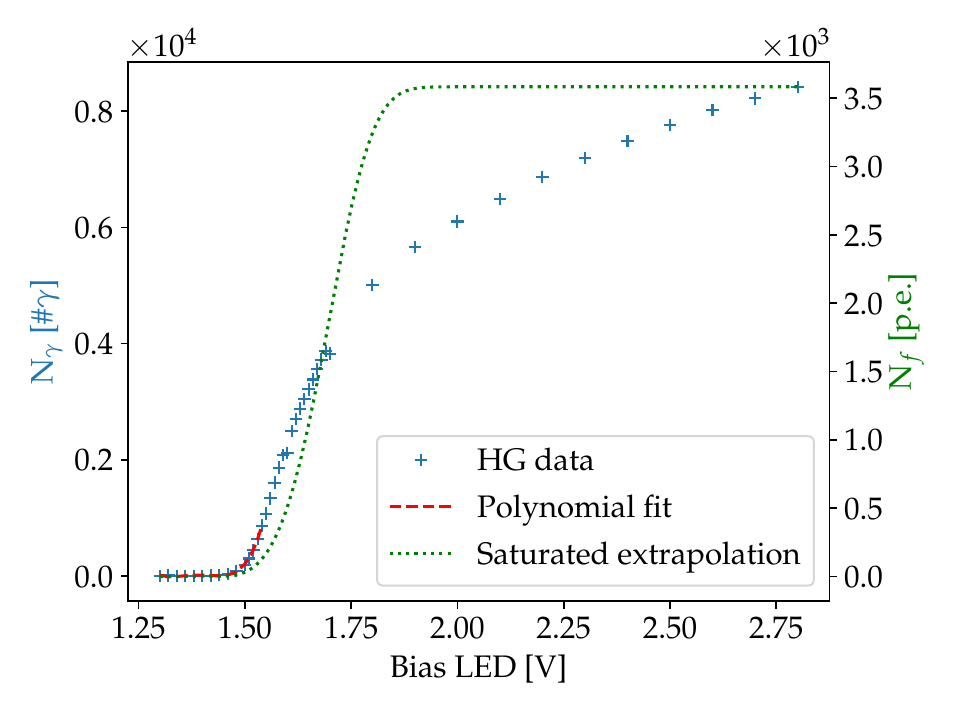}
  \par\end{centering}
  \protect\caption{\textbf{Left:} Amplitude and charge of the HG pulse as a function of the LED bias voltage~\cite{key-thesisNDA}. \textbf{Right:} 4-degrees polynomial fit\protect\footnotemark ~of the non-saturated charge as a function of the LED bias voltage and extrapolation of the fit combined with the geometrical saturation formula \eqref{eq:geom_satu}.}
    \label{fig:amp_charge_nocalib_calibLED}
\end{figure}
\par\end{center}
\footnotetext{The polynomial fit parameters are: $p_4=(1.95\pm0.21)\cdot10^5$~V$^{-4}$, $p_3=(-1.08\pm0.12)\cdot10^7$~V$^{-3}$, $p_2=(2.24\pm0.26)\cdot10^7$~V$^{-2}$, $p_1=(-2.07\pm0.24)\cdot10^7$~V$^{-1}$, and $p_0=(7.15\pm0.86)\cdot10^6$.}

The relationship between the bias voltage of the LED and the real light level is determined by calibrating the LED. 
The gain for both amplitude and charge is determined by fitting the multi-photoelectron spectrum, which allows conversion of amplitude and charge into a number of photoelectrons corrected for optical crosstalk. 
The gains obtained from the spectra are 112.08~mV$\cdot$ns for the charge and 3.4771~mV for the amplitude. Combining these gains with the curves shown in Figure~\ref{fig:amp_charge_nocalib_calibLED} leads to the number of photoelectrons versus the LED bias voltage for both charge and amplitude. 
The curve obtained for the charge is shown in Figure~\ref{fig:amp_charge_nocalib_calibLED}-left, where the number of photoelectrons has been converted to a number of photons by accounting for the 36\% PDE (at $\lambda$\textsubscript{LED}=525~nm) and for the 8\% optical crosstalk. 
The non-saturated part of this curve, which ranges up to 1.55~V, is fitted using a 4-degree polynomial function (which reproduces the LED behavior) in order to extract a relation between the number of photons emitted by the LED and its bias voltage.

To account for the geometrical saturation due to the limited number of micro-cells in the sensor ($N_{cells}$=3584), the 4-degree polynomial fit has to be convoluted with relation \eqref{eq:geom_satu}, which expresses the number of fired cells $N_f$ as a function of the number of photons $N_\gamma$ \cite{D_Renker_2009, Barral2004}:

\begin{equation}
\label{eq:geom_satu}
N_f = N_{cells} \qty[1-\exp(-\frac{N_\gamma \cdot PDE \cdot (1+\mu)}{N_{cells}})]
\end{equation}

where $N_{cells}$ is the number of micro-cells, PDE is the photodetection efficiency at the LED wavelength, and $\mu$ is the crosstalk probability.

%\vspace{0.4cm}
The resulting relation between the number of fired micro-cells and the bias voltage of the LED is plotted in Figure~\ref{fig:amp_charge_nocalib_calibLED}. 
This allows converting the x-scale of the initial plots (Figure~\ref{fig:amp_charge_nocalib_calibLED}) into a number of fired micro-cells. 
This is shown in Figure~\ref{fig:scan_geom_satu} for both amplitude and charge. We can see a good linear behavior until about 300 fired micro-cells for the amplitude. The charge has also a good behavior in saturation which extends the dynamic range up to 3000 photoelectrons and demonstrate that a single gain (HG) is enough to cover the required dynamic range.

\begin{center}
\begin{figure}[H]
\captionsetup{format=myformat}
\begin{centering}
\hspace*{-.5cm}\includegraphics[height=.4\textwidth]{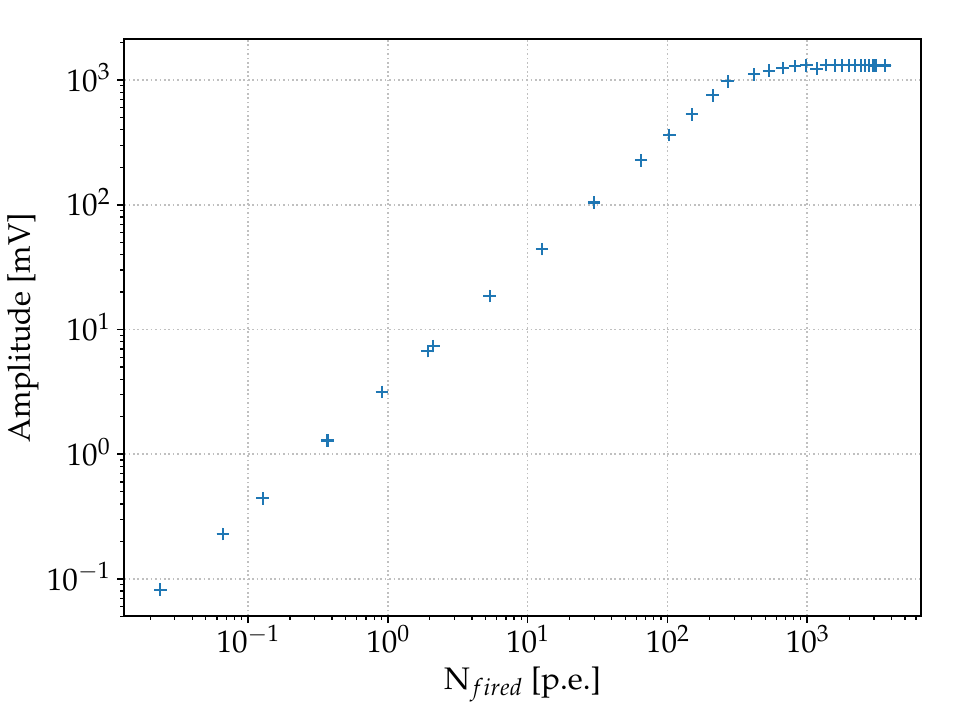}\includegraphics[height=.4\textwidth]{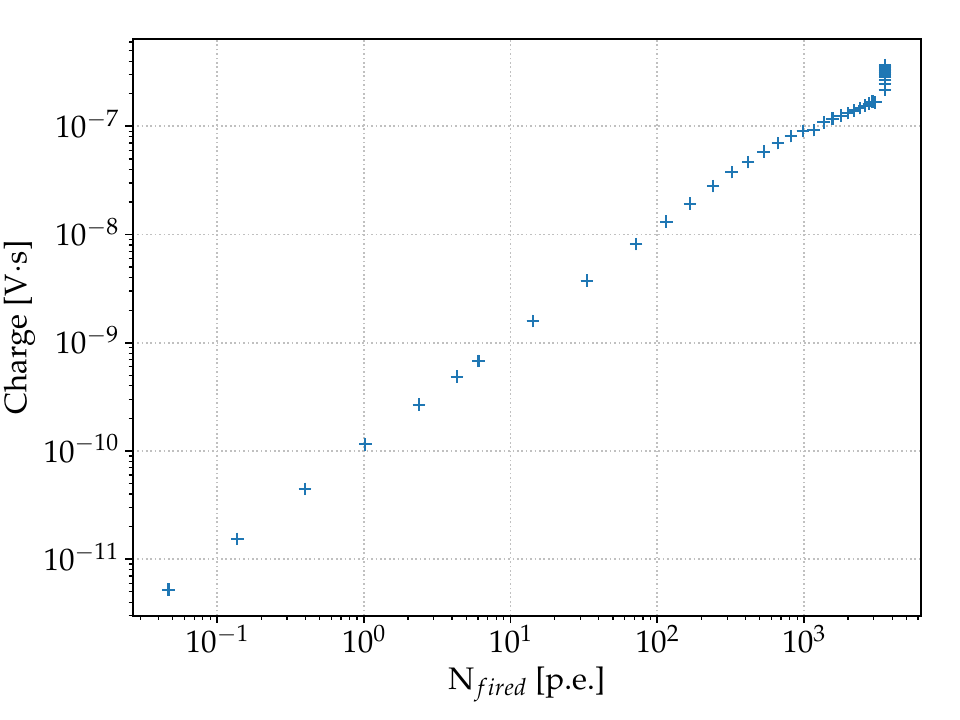}
  \par\end{centering}
  \protect\caption{Amplitude and charge of the HG pulse as a function\newline of the light level accounting for the geometrical saturation \cite{key-thesisNDA}}
    \label{fig:scan_geom_satu}
\end{figure}
\par\end{center}

The same linearity measurements have been performed for the LCT5 3$\times$3~mm$^2$ sensor with the intermediate PZC configuration (R=4, C=9) and same the behavior is found for both cases.

\section{Comparison between simulations and measurements using LCT5 sensors coupled to the MUSIC ASIC}

Some experimental measurements with the SiPMs and the MUSIC ASIC have been performed and compared to simulations, as shown in Figure~\ref{fig:music_simu_shape_lct5}.
In this case, the MUSIC ASIC is configured in the individual analog readout mode, i.e., only a single channel is evaluated. Although, the summation response when reading a single channel behaves similarly. Moreover, the PZC circuit has been tuned with different configurations or bypassed (indicated with PZC off).  
The resistance R and capacitance C controls an RC filter to compensate for the decay time of the SiPM. As can be seen from Figure~\ref{fig:music_simu_shape_lct5}, the larger the resistance, the narrower becomes the SiPM response at the expense of larger attenuation of the signal and the consequent loss in SNR. 

It can be seen that simulations and measurements are quite similar, especially, for the cases when the PZC is applied; the divergence in some cases might be due to inaccuracies in the simulation with respect to the experimental test bench.
For instance, differences may arise from larger parasitic components in the manufactured chip compared with the expected value, packaging process, PCB design or cables.

\begin{center}
\begin{figure}[H]
\captionsetup{format=myformat}
\begin{centering}
\hspace*{-1cm}\includegraphics[height=.4\textwidth]{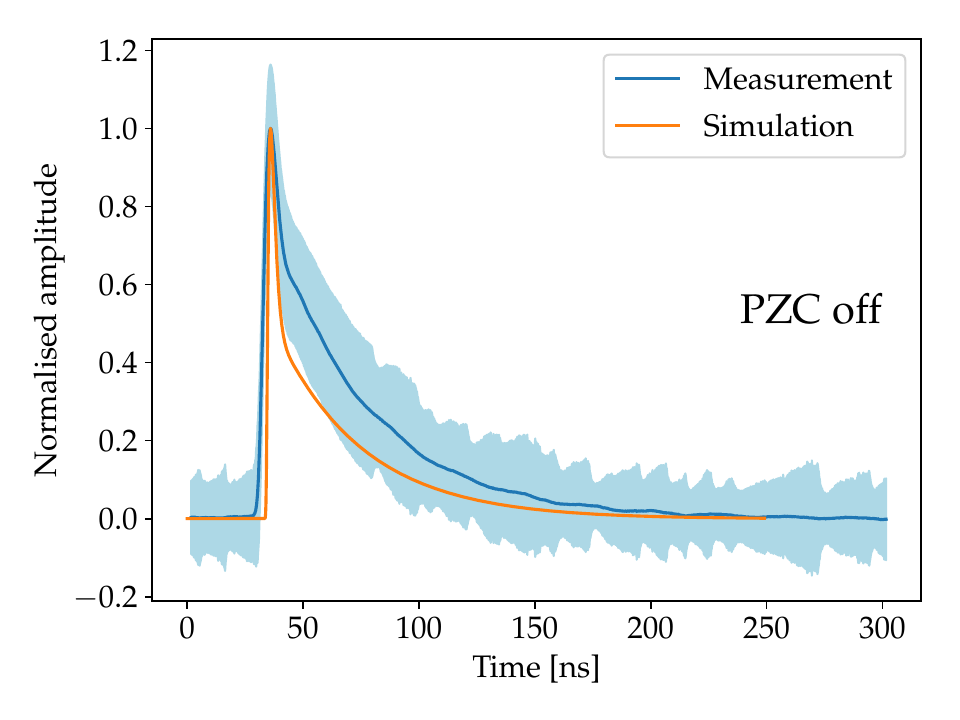}\includegraphics[height=.4\textwidth]{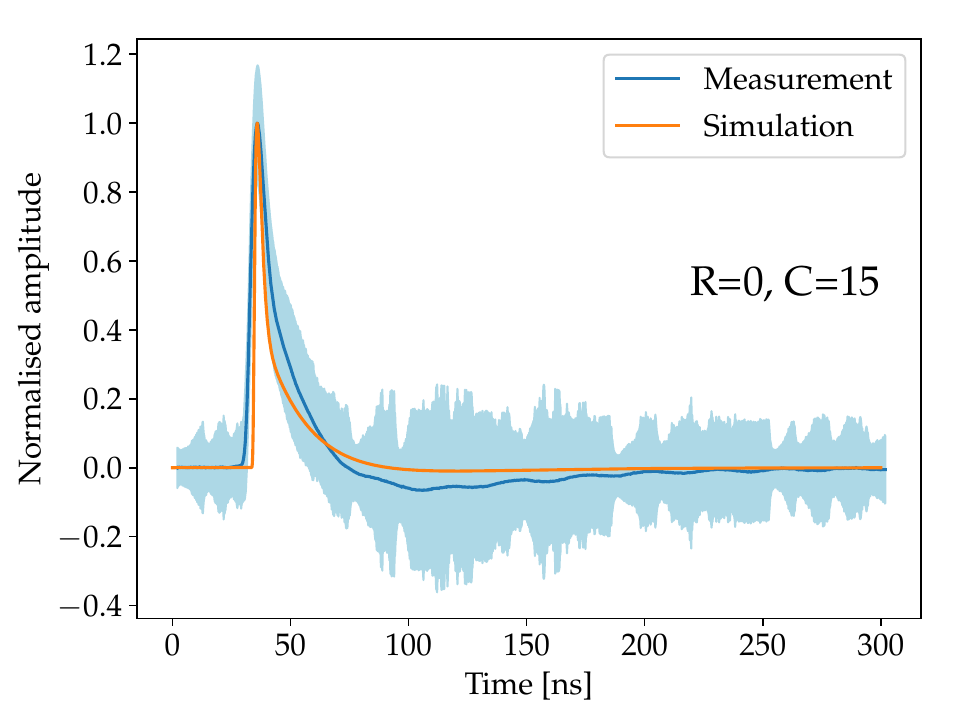}
\noindent\hspace*{-1cm}\includegraphics[height=.4\textwidth]{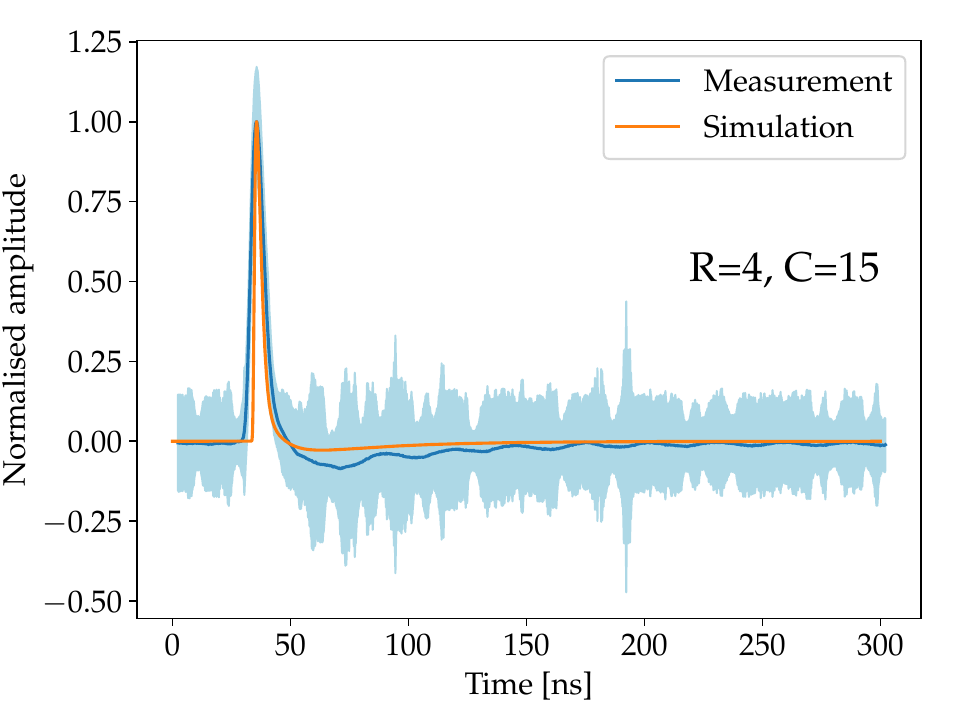}\includegraphics[height=.4\textwidth]{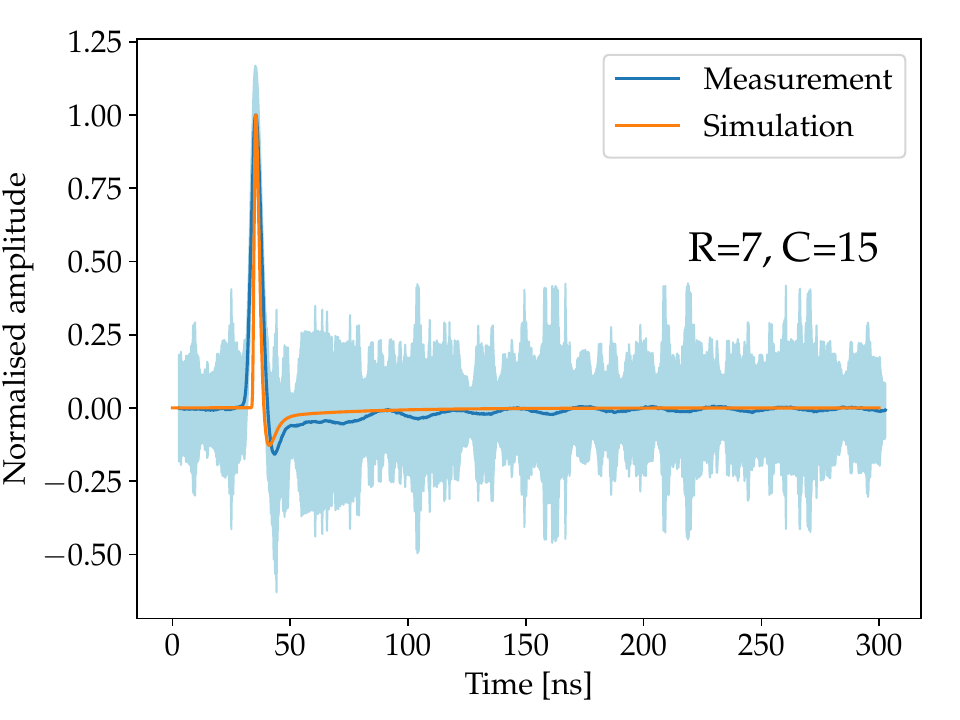}
  \par\end{centering}
  \protect\caption{LCT5 3$\times$3~mm\textsuperscript{2} pulse shapes comparison between measurements\newline and simulations for different PZC configurations (adapted from \cite{key-thesisNDA}).}
    \label{fig:music_simu_shape_lct5}
\end{figure}
\par\end{center}

\label{sec-hexa}
\section{Response prediction of a hexagonal sensor with LCT5 technology}

In~\cite{key-Sensors}, a SiPM with a hexagonal-like shape was proposed using HPK technology, referred to as S10943-2832(X) (LCT2). We intend to use the same sensor shape for the upgraded camera, hence it is relevant to simulate the response of the readout electronics to a LCT5 sensor of such shape.

In this case study, four LCT5 6$\times$6~mm$^2$ (S13360-6050CS) have been used to predict the performance of a SiPM with an hexagonal shape using the sum output of the MUSIC ASIC. For the sake of completeness, the same was performed with the LVR3 6$\times$6~mm$^2$ (S14520-6050VS) technology.
The number of micro-cells in the S10943-2832(X) has been used to simulate the LCT5 and LVR3 hexagonal sensors. Moreover, a standard square S13360-6050CS has been also used to evaluate the performance of an existing sensor. 

Simulations have been performed by injecting the signal coming from four identical 6$\times$6~mm$^2$ SiPMs into the MUSIC circuit when PZC  (R=4, C=15) is active. Table~\ref{tab:snr} shows the SNR that can be achieved using these four types of SiPMs in terms of peak amplitude and after integrating the first 10~ns of the pulse.
%Only this fraction of signal has been used for integration since it contains the significant information of the response. Note that larger the integration time, larger the probability to suffer pile-up of dark and NSB pulses, cross-talk and after pulses. 
As it can be seen from Table~\ref{tab:snr}, all hexagonal sensors have an equivalent capacitance smaller than the standard square SiPM and thus electronics noise is lower for these sensors. 
%As said in introduction, the larger is the SiPM capacitance, the larger the electronic noise~\cite{key-MUSIC_journal} and smaller the peak amplitude~\cite{key-simus-paper} are. 
As said in the introduction, a large capacitance in general leads to a small SNR. It should be noted that the SNR for the charge measurement  does not always decrease with capacitance, since it also depends on how the charge is spread over the signal tail, i.e., how fast the SiPM is discharged. Lastly, as expected, the SNR is smaller when considering the peak amplitude instead of the charge. 
When using the amplitude, we are just using a single point of the acquired waveform which can be more susceptible to other noise sources such as ringing from inductance. 
By using the charge method, as the integral of the signal where most of the information is contained, we are using more information from the signal to evaluate the response and thus we are less sensitive to other noise fluctuations such as the aforementioned ringing.

% The large detection area needed for the future LST can be divided in a cluster of 4 pixels. HPK proposed this SiPM in the past. 
% In this study, we imitated this SiPM-shape but using newer SiPM models.
% Peaks can be clearly identified when summing 1 SiPMs.
% PZ configured is R=4, C= 15.
% SNR (charge) is obtained by integrating the output of the MUSIC chip during 10~ns.
% Gain music: 480 HG

\begin{table}[ht]
    \small
	\centering
	\renewcommand*{\arraystretch}{1.4}
	\setlength\arrayrulewidth{1.5pt}
	\begin{tabular}{|c|c|c|c|c|c|}
		\hline
		\rowcolor{flavescent} 
		\textbf{Parameter}  & \textbf{S10943-2832X} & \textbf{S13360-XX50} & \textbf{S14520-XX50} & \textbf{S13360-6050CS} \\
		\rowcolor{flavescent} 
		\textbf{}  & \textbf{Hexagonal} & \textbf{Hexagonal} & \textbf{Hexagonal} & \textbf{} \\
		\hline
		\rowcolor{lightcornflowerblue}
		Number of Pixels & 9210  & 9210  & 9210 & 14400  \\	 
		\hline
		\rowcolor{lavendergray} 	
		Capacitance [pF] & 851  & 944 & 1325 & 1477  \\  
		\hline
		\rowcolor{lightcornflowerblue}
		SNR (Peak amplitude) & 6.6  & 5.3  & 4.8 & 3.7  \\	 
		\hline
		\rowcolor{lavendergray} 	
		SNR (Charge) & 10.8  & 6.0 & 4.6 & 5.7  \\  
		\hline
	\end{tabular}
	\caption{Simulated SNR (expressed in linear scale) using the peak amplitude and the charge over 10~ns integration windows for different proposed sensors. Over-voltage is set to 8~V.}

	\label{tab:snr}
	%\vspace*{-0.2cm} 
\end{table}

Experimental measurements have been carried out connecting the MUSIC to four square standard S13360-6050CS SiPMs (LCT5 6$\times$6~mm$^2$). 
The ASIC is configured in summation mode with PZC (R=4, C=15). This represents a signal with a FWHM of the order of 10~ns. 
Figure~\ref{fig:music_amp_osc} shows the analog response of the SiPM when illuminating the sensors at the single photon regime.
At 6~V over-voltage, the different photons can be clearly distinguished in the peak amplitude histogram.
%can be slightly identified in the color representation, but  when representing the peak amplitude histogram the peaks can be distinguished.
As expected from simulations, when increasing the over-voltage to 8~V the peaks can be better identified at the expense of larger correlated and non-correlated SiPM noise. 
Figure~\ref{fig:music_charge_lct5} shows the charge histogram when integrating the output pulse over 10~ns. The single photo-electron response can be already identified at 4~V over-voltage, and at 8~V the peaks are better separated, as expected. 
%Moreover, a linear behavior of the summation response is observed, as expected in~\cite{key-MUSIC_journal}. 
To quantify this statement, the SNR can be measured. It is defined as the ratio of the gain (measured as the distance in charge between the first peak (pedestal) and the second peak, i.e., the 1\textsuperscript{st} photo-electron) and the standard deviation of the first peak (electronic noise). Using the charge, we find two values of the SNR of 3.2 and 5.6 for 4~V and 8~V over-voltage, respectively. The SNR for 8~V agrees with the simulation, as shown in Table~\ref{tab:snr}.

\begin{center}
\begin{figure}[H]
\captionsetup{format=myformat}
\begin{centering}
\includegraphics[height=.3\textwidth]{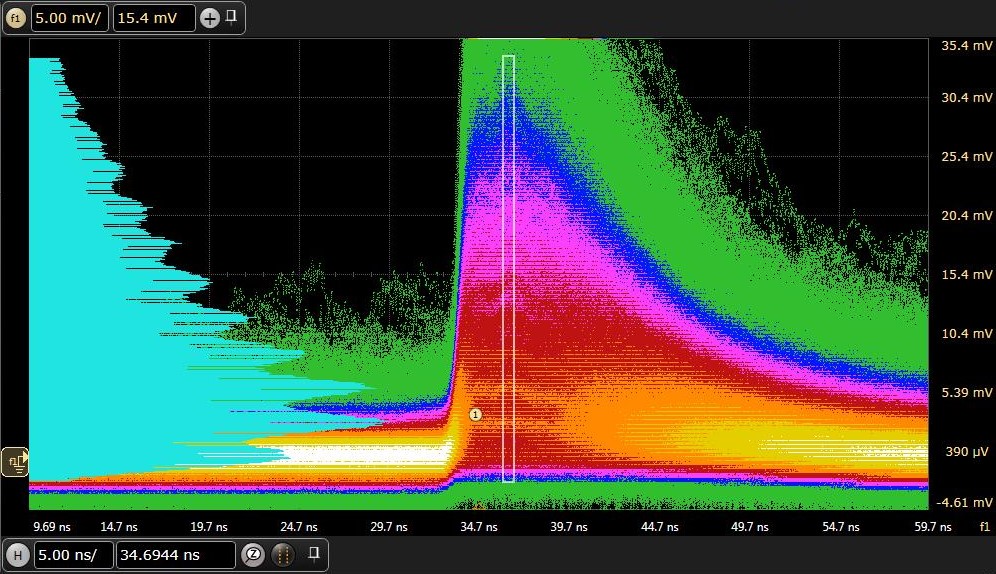}\includegraphics[height=.3\textwidth]{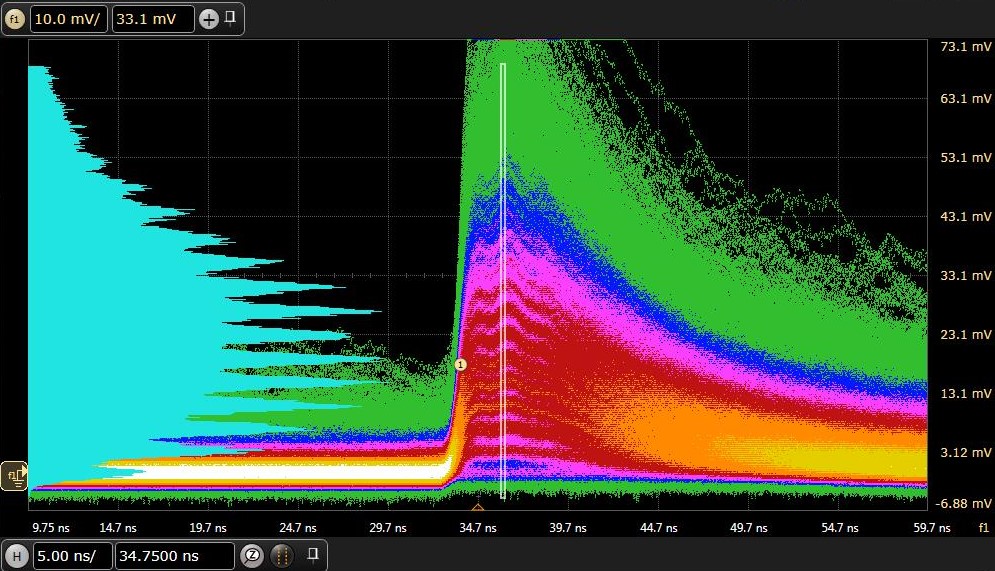} 
  \par\end{centering}
  \protect\caption{Response of the MUSIC connected to four S13360-6050CS SiPMs in summation mode and with PZC (R=4, C=15). The histogram is based on the highlighted peak of the summation output. (left) 6~V of over-voltage and (right) 8~V of over-voltage.}
  \label{fig:music_amp_osc}
\end{figure}
\par\end{center}

\begin{center}
\begin{figure}[H]
\captionsetup{format=myformat}
\begin{centering}
\includegraphics[height=.4\textwidth]{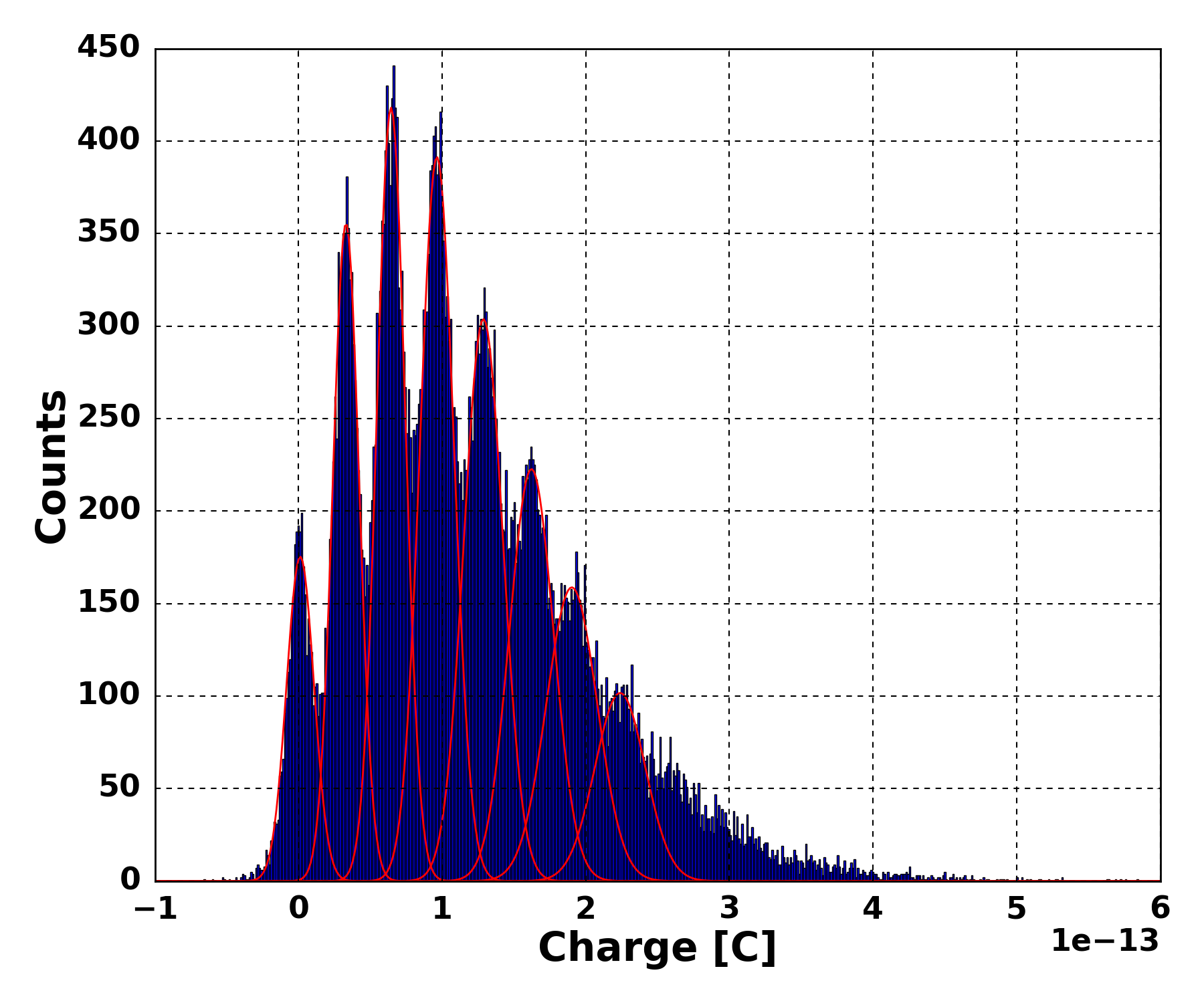}%\includegraphics[height=.4\textwidth]{Pictures/charge_by_cell-charge_dist-waveform_4_57.4_MUSIC1_S13360-6050CS_50K_SUM_numCh1.png}
\includegraphics[height=.4\textwidth]{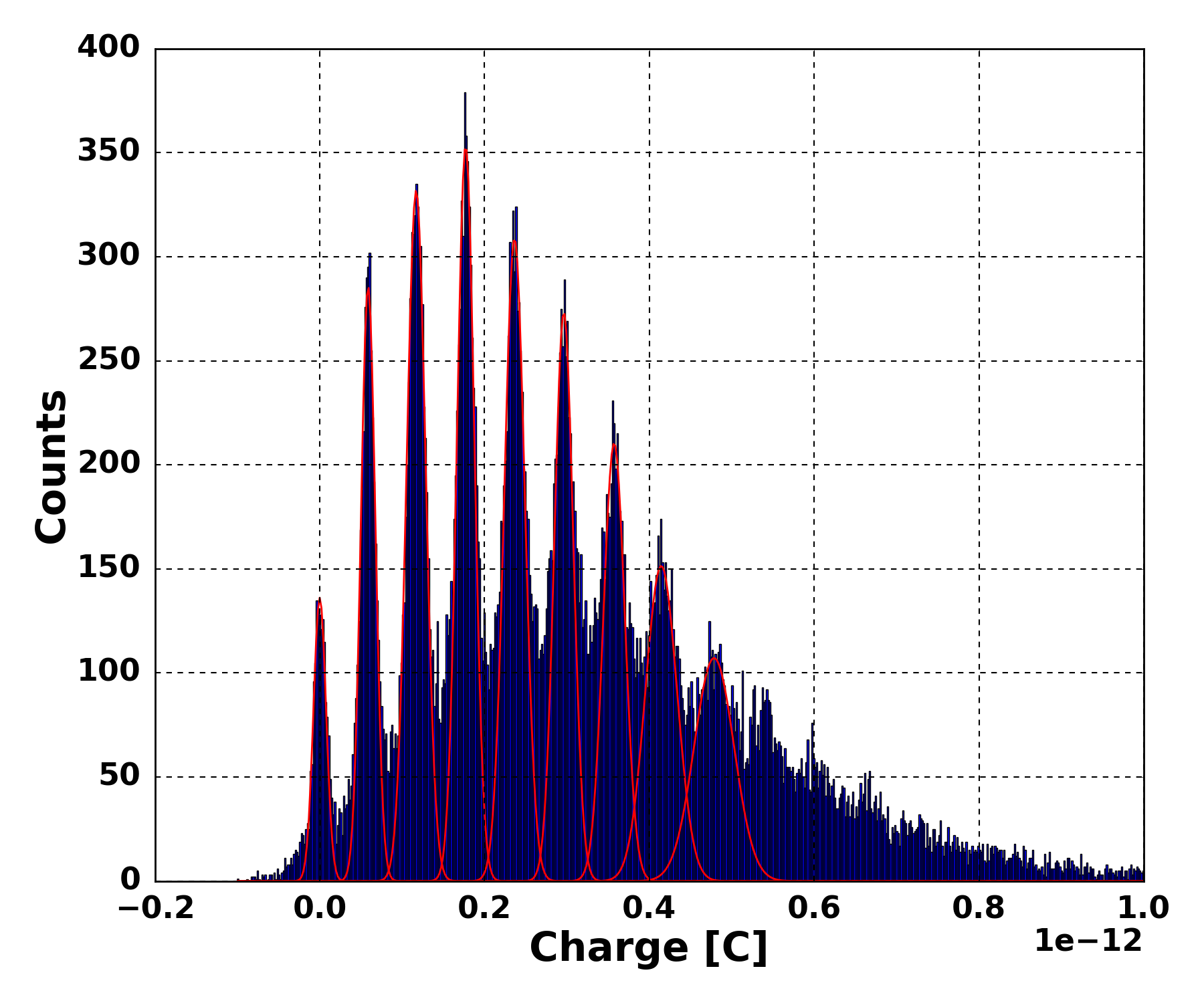}%\includegraphics[height=.4\textwidth]{Pictures/charge_by_cell-charge_dist-waveform_4_61.4_MUSIC1_S13360-6050CS_50K_SUM_numCh1.png}
  \par\end{centering}
  \protect\caption{Charge histogram of the MUSIC connected to four S13360-6050CS SiPMs in summation mode and with PZC (R=4, C=15). (Left) 4~V of over-voltage; and (Right) 8~V of over-voltage.}
    \label{fig:music_charge_lct5}
\end{figure}
\par\end{center}

%\newpage
\section{Conclusions and discussion}

% Conclusions

This work presents a method to foresee the performance of a large area sensor combined to a preamplifying electronics based on the simple model for SiPM, the Corsi model~\cite{CORSI2007416, Corsi}. Particularly for our case, where we developed a dedicated hexagonal sensor which requires a dedicated photo-mask by a company, this becomes very useful as one can exactly plan the required performance and then implement the sensor according to such design.

In order to anticipate the response of the pre-amplifying stage, the model of the sensor and the integrated circuit used for its readout have been validated on measurements. The excellent agreement between measured and simulated pulses has allowed anticipating with confidence the response of the MUSIC ASIC to hexagonal sensors produced in the LCT5 and LVR3 technology. Based on this work, predictions can be made for any custom SiPM. It was shown that the Corsi model parameters can be carefully extrapolated from an existing smaller sensor. 
The properties of the available standard and smaller S13360-6050CS SiPMs of 6$\times$6~mm$^2$ have been used to experimentally prove that the 1~cm$^2$ hexagonal sensors obtained with LCT5 and LVR3 technologies have  photo-electron identification capabilities. The MUSIC ASIC is perfectly capable of treating large sensors built out of smaller units thanks to its sum output. For the tested sensor, it produces an output pulse with approximately 10~ns FWHM and a power consumption of about 100~mW. 

If this performance is enough for small telescopes which work in the highest energy range, further improvements must be achieved to reach pulse widths as low as 3~ns FWHM. This will help improving performance in the lowest energy range of the large-sized telescopes of CTA. 
%If some research and development is carried out by the SiPM manufacturers to reduce the recovery time of the sensors, 
A future ASIC, based on the 8-channel FastIC ASIC developed in a 65~nm technology node~\cite{Fastic_2022,FASTIC_NSS_2021} is planned for the future upgrade of the LSTs cameras using SiPMs. 
This new ASIC 
%will be developed in 65~nm and 
should have 16 input channels, the capability to perform an active summation in groups of four channels, the option to add the signals from these 16 channels into a single output, a mechanism to capture the level of continuous night sky background (NSB) noise and the possibility to drive cable loads. The power consumption of the ASIC should be around 10~mW/ch. As a second version of the chip, an ADC is planned to be added in order to provide a digital output directly from the ASIC.

\section*{Acknowledgements}

We acknowledge the support of the University of Geneva, the Swiss National Foundation FLARE Program, grant 20FL21\_201539 and of the State Secretariat for Education Research and Innovation on CTAO. N.D.A. acknowledges the support of the Swiss National Science Foundation. We acknowledge support by the Spanish Ministerio de Ciencia, Innovación y Universidades 491 (MICINN), grant FPA2017-82729-C6-2-R (AEI/FEDER, UE) and grant PID2019-104114RB-C33 / AEI / 10.13039/501100011033. We also acknowledge financial support from the State Agency for Research of the Spanish Ministry of Science and Innovation through the “Unit of Excellence María de Maeztu 2020-2023” award to the Institute of Cosmos Sciences (CEX2019-000918-M). 

\normalem
\printbibliography

\end{document}